\documentclass[11pt,aps,amsfonts,nofootinbib,preprintnumbers,nobalancelastpage,showkeys]{revtex4-1}
\usepackage[english]{babel}
\usepackage{amsmath,amssymb}
\usepackage{amsfonts}
\usepackage[dvips]{graphicx}
\usepackage[sort&compress]{natbib}
\usepackage{bbm}
\usepackage{subfigure}
\usepackage{color}
\usepackage{footnote}
\usepackage{slashed} 
\usepackage{multirow}
\usepackage{braket}

\makeatletter
\def\endfmffile{%
  \fmfcmd{\p@rcent\space the end.^^J%
          end.^^J%
          endinput;}%
  \if@fmfio
    \immediate\closeout\@outfmf
  \fi
  \IfFileExists{\thefmffile.mp}{\immediate\write18{mpost \thefmffile}}{}
  \let\thefmffile\relax
}
\makeatother
\usepackage[normalem]{ulem}

\newcommand{\ie} {{\it i.e.}}

\newcommand {\beq} {\begin{equation}}
\newcommand {\eeq} {\end{equation}}
\newcommand {\bea} {\begin{eqnarray}}
\newcommand {\eea} {\end{eqnarray}}

\newcommand{\maddm}{MadDM}
\newcommand{\madgraph}{MadGraph5$\underline{\hspace{0.16cm} }$aMC@NLO{}}

\definecolor{red1}{cmyk}{0,1,1,0.1}
\definecolor{blue1}{cmyk}{1,0,0,0}

\newcommand{\ignore}[1]{}
\newcommand{\nn}{\nonumber} \renewcommand{\bf}{\textbf}
\newcommand{\mbf}{\mathbf}

\newcommand{\GeV}{{\rm\ GeV}}

\newcommand{\kg}{{\rm\ kg}}

\newcommand{\esc}{{\rm esc}}

\allowdisplaybreaks

\keywords{Beyond Standard Model, MadGraph, Dark Matter, Relic Density, Numerical Tools, Directional Dark Matter, LHC}

\begin{document}

\preprint{{\small CP3-15-13, MCNET-15-10}}

\title{Direct Detection of Dark Matter with MadDM  v.2.0}
\date{\today}
\author{Mihailo Backovi\'{c}} \email{mihailo.backovic@uclouvain.be} \affiliation{Centre for Cosmology, Particle Physics and Phenomenology (CP3), \\ Universit\'{e} catholique de Louvain,  Chemin du Cyclotron 2, \\ B-1348 Louvain-la-Neuve, Belgium} 
\author{Antony Martini} \email{antony.martini@uclouvain.be} \affiliation{Centre for Cosmology, Particle Physics and Phenomenology (CP3), \\ Universit\'{e} catholique de Louvain,  Chemin du Cyclotron 2, \\ B-1348 Louvain-la-Neuve, Belgium} 
\author{Olivier Mattelaer}\email{o.p.c.mattelaer@durham.ac.uk}\affiliation{Institute for Particle Physics Phenomenology (IPPP), Durham University, Durham, DH1 3LF, United Kingdom}
\author{Kyoungchul Kong} \email{kckong@ku.edu} \affiliation{Department of Physics and Astronomy, \\ University of Kansas, Lawrence, KS, 66045, USA} 
\author{Gopolang Mohlabeng} \email{gopolang.mohlabeng@ku.edu} \affiliation{Department of Physics and Astronomy, \\ University of Kansas, Lawrence, KS, 66045, USA} 

\begin{abstract}
\vspace*{1cm}
We present \maddm{} v.2.0, a numerical tool for dark matter physics in a generic model. This version is the next step towards the development of a fully automated framework for dark matter searches at the interface of collider physics, astro-physics and cosmology. It extends the capabilities of v.1.0 to perform calculations relevant to the direct detection of dark matter. These include calculations of spin-independent/spin-dependent nucleon scattering cross sections and nuclear recoil rates (as a function of both energy and angle), as well as a simplified functionality to compare the model points with existing constraints. The functionality of \maddm{} v.2.0 incorporates a large selection of dark matter detector materials and sizes, and simulates detector effects on the nuclear recoil signals. 
We validate the code in a wide range of dark matter models by comparing results from \maddm{} v.2.0 to the existing tools and literature.  

\end{abstract}
\maketitle

\tableofcontents

\newpage
\section{Introduction}\label{sec:intro}

While there are many intriguing, indirect hints of the existence of dark matter (DM), the evidence of irrefutable signals and the true nature of DM remain elusive 
(see Ref. \cite{Bergstrom:2012fi} and references therein for recent progress in DM searches).
Various direct detection experiments are currently ongoing, and many are under development. The results from direction detection experiments  have so far produced stringent constraints on the DM scattering cross section off atomic nuclei, especially  for the DM in the mass range of the electroweak (EW) scale.
More recently, the LHC analyses of jet(s) + missing energy and $W/Z/t$ + missing energy channels, and their variants, provided complementary information to underground fixed target experiments. The complementarity was particularly evident in the improved sensitivity for a relatively light DM in simplified model scenarios \cite{Abdallah:2014hon}. 
Despite a significant effort, none of the individual experiments have provided a strong indication on the mass scale of dark matter particles.
It is therefore essential to search for evidence of dark matter in many different ways: 
by directly producing DM candidate at particle colliders, as well as by detecting it in our galaxy and beyond \cite{Arrenberg:2013rzp}. 

Searching for DM at the interface of collider physics, astroparticle physics and cosmology is perhaps the most promising path to discovery and discrimination of different DM models, which requires substantial amount of computational power and relevant tools. 
While many collider event generators are publicly available \cite{Ask:2012sm}, the number of numerical tools which are able to calculate signals of dark matter at the interface of collider physics and astroparticle physics remains very limited \cite{Gondolo:2004sc,Belanger:2014hqa}. \maddm~\cite{Backovic:2013dpa} emerged as an attempt to build such a tool. The goal of the \maddm{} project is to form a simple, user friendly and all-in-one DM phenomenology framework. Our ambition is to allow both experimentalists and theorists to calculate accurate signatures of generic DM models at colliders, in our galaxy and in the early universe with only a few clicks of a button. 
We also aim to provide flexibility such that users easily replace the existing modules with their own implementation or link \maddm{} functionality with other existing tools.

The user friendly architecture of the \madgraph{} \cite{Alwall:2011uj,Alwall:2014hca} package provides an ideal framework for the development of MadDM. Numerous MadGraph additions and extensions offer many interesting directions that development of \maddm{} can go into. Version 1.0 of \maddm{} focused on computation of DM relic density in a generic model \cite{Backovic:2013dpa}. 
Here we discuss the next step in the development of MadDM: direct detection of galactic DM. 
In addition to the ability to calculate DM-nucleon cross sections, \maddm{} v.2.0 goes a step further and provides the directional information of nuclear recoil, as well as the ability to simulate the effects of detector systematics on recoil distributions. 
To the best of our knowledge, \maddm{} is the first publicly available code with such a capability. 
The code is able to calculate double differential recoil rates, as a function of recoil angle and energy, 
as well as the energy distribution and angular distribution of nuclear recoil.

Directional information of DM scattering can be of great importance for the next generation dark matter direct detection experiments. 
In case a DM signal is observed, directional information about nuclear recoil rates could lead to more precise measurements of intrinsic particle properties of DM \cite{Mohlabeng:2015efa} as well as astrophysical information such as the DM velocity distribution \cite{Copi:1999pw,Gondolo:2002np}. Conversely, if next generation direct detection experiments lead only to more stringent limits on DM mass and scattering cross section, directional detection could become crucial in overcoming the irreducible neutrino background \cite{Grothaus:2014hja}.  

Finally, \maddm{} v.2.0 incorporates a simplified functionality for testing model points against experimental constraints, whereby the user can choose the range of DM relic densities and upper bounds on DM-nucleon scattering cross sections for which the model is consistent with the experiments. The code will automatically compare the output of the calculation to the user specified constraints and determine whether the model point is allowed or not.

New version of \maddm{} follows the already established structure of \maddm{} v.1.0. A \verb|Python| module generates relevant amplitudes for relic density and direct detection (with ALOHA \cite{deAquino:2011ub}), while a \verb|FORTRAN| module handles the heavy numerical calculations. As before, \maddm{} is compatible with UFO 
(Universal FeynRules Output \cite{Degrande:2011ua}) model which contains a dark matter candidate and can be easily linked to any width or mass spectrum generator which can produce a Les Houches formatted parameter card \cite{Skands:2003cj,Allanach:2008qq}. 

The following sections describe the new functionality of the \maddm{} code for the direct detection of DM, while we refer the reader to Ref. \cite{Backovic:2013dpa} for more information on calculation of DM relic density. Section \ref{sec:review} describes the theoretical background of dark matter direct detection, 
including directionality. 
A description of the new routines in the \maddm{} code is presented in Section \ref{sec:code}. Finally, we show several validations and example calculations using the \maddm{} v.2.0 code in Section \ref{sec:validations}. We reserve the appendix for a brief description of the \maddm{} code structure.


\section{Review of Direct Detection and its Implementation in MadDM}\label{sec:review}
\subsection{DM-nucleon Elastic Cross Section}\label{sec:DirDetect}
The possibility that DM could weakly interact with ordinary matter has led to a development of a myriad of experiments which search for signatures of galactic DM scattering off nuclei \cite{Faham:2014hza,Brown:2014bda,Saab:2014lda,PhysRevLett.107.141301,Bernabei:2010mq}.
Being sensitive to the nuclear recoil of the scattering DM-nucleon event in underground detectors, experiments can constrain model parameters involved in the DM-nucleon (nucleus) elastic scattering cross section, while in the case of a discovery, the same information may be used to determine properties of DM particles. The abundance of independent experiments has so far led to important limits on the DM-nucleus cross-section, 
among which the most recent (and most stringent) one comes from the LUX collaboration \cite{Akerib:2013tjd,Faham:2014hza}.
In this section, we give an overview of the theoretical formalism of DM-nucleus elastic scattering 
in order to allow the reader to understand the procedures \maddm{} performs in computing DM direct detection rates \cite{Dobrescu:2007ec,Freytsis:2010ne,Belanger:2008sj}.  

 The \maddm{} v.2.0 code incorporates the following definition of the DM-nucleus \\ \textbf{spin-independent} (SI) scattering cross section:
  \beq
    \sigma_{SI} = \frac{4}{\pi}\mu_A^2\cdot \left[ Z \cdot f_p + \left( A-Z \right) \cdot f_n \right]^2 \, ,    \label{SIcx}
  \eeq
  where $\mu_A=\frac{m_\chi m_A}{m_\chi+m_A}$ is the DM-nucleus reduced mass and $f_p$ and $f_n$ are proton/neutron spin-independent form factors respectively \cite{Grothaus:2014hja,Spergel:1987kx,Copi:1999pw}. 
  We use $Z$ to denote the number of protons in a nucleus and $A$ to denote the number of protons and neutrons.
  Similarly, for \textbf{spin-dependent} (SD) \textbf{interactions}, we use the definition: 
  \beq
    \sigma_{SD} = \frac{16}{\pi} \mu_A^2\cdot \frac{J_A + 1}{J_A}\left(f'_p+ f'_n\right)^2,
    \label{SDcx}
  \eeq
  where $f'_p$ and $f'_n$ are the proton/neutron spin-dependent form factors respectively, and $J_A$ is the spin of the nucleus. 
Particle physics enters the calculation of $\sigma_{SD/SI}$ via the form factors $f_N$ and $f'_N$ where $N = p,n$.

The spin-independent quantity for a scalar current, $f_N^S$ is defined as
\beq
	 f_N^S= m_N \sum_{q=u,d,s} \frac{\alpha_q}{m_q} f_N^{Sq} + \frac{2}{27} m_N f_N^{SG} \sum_{q=c,b,t} \frac{\alpha_q}{m_q},
\eeq
where $\alpha_q$ are the coefficients of the low energy matrix elements defined by 
\beq
\Braket{{\mathcal{M}}} (\chi q \rightarrow \chi q) = \alpha_q \Braket{\bar{\psi}_q \psi_q} \, .
\eeq
The quantities $f_N^{S\, q/G}$ are nucleon form factors related to the low energy elastic scattering matrix elements as follows
 \begin{align}
   \Braket{N \left| \bar{\psi}_q \psi_q \right| N} &= ~\frac{m_N}{m_q} \cdot f_N^{Sq}  \quad~~ ~~ \mbox{(for $u$, $d$, $s$)} \, , \\
   \Braket{N \left| \bar{\psi}_q \psi_q \right| N} &= ~ \frac{2}{27} \frac{m_N}{m_q} \cdot f_N^{SG} \quad  \mbox{(for $c$, $b$, $t$)} \, .
 \end{align}
The  values for nucleon form factors $f_N^S$ are typically extracted from data or from chiral perturbation theory \cite{Beringer:1900zz} and depend on the type of current which defines the DM-nucleon interaction.  The default values for \textbf{scalar} spin-independent form factors of light quarks used in \maddm{} are:
\begin{eqnarray}
f_p^{Su} = 0.0153,  \,\,\,\,\,& f_n^{Su} = 0.0110, \nn \\
f_p^{Sd} = 0.0191, \,\,\,\,\,& f_n^{Sd} = 0.0273,  \\
f_p^{Ss} = 0.0447, \,\,\,\,\, & f_n^{Ss} = 0.0447,  \nn
\end{eqnarray}
while the gluon scalar form factor is
\begin{equation}
f_N^{SG} = 1- \sum_q {f_N^{Sq}} \, .
\end{equation}

The \textbf{vector current} interaction is characterized by a simpler set of form factors. Since the vector current is conserved, the DM-nucleon form factor can be obtained from  the sum of the currents of the valence quarks, \ie:
  \beq
    f_N^V = \sum_{q=u,d} f_N^{Vq} \cdot \alpha_{q}\,,
  \eeq
where $f_p^{Vu}  = f_n^{Vd} = 2$ and $f_p^{Vd} = f_n^{Vu} =  1$. \\

For spin-dependent interactions, the following equation defines the proton/neutron form factors $f_N^{'}$ in Eq.~\eqref{SDcx}:
\beq
 f_N^{'} = \sum_{q=u,d,s} \alpha_{q} \, \Delta_N^{q} \, ,
 \label{SDfprime}
\eeq
where the quantities $\Delta_N^q$ are nucleon form factors, applies for both \textbf{axial-vector} and \textbf{tensor currents}. In {\maddm{}}, we use the following default values for the axial-vector interaction:
\begin{eqnarray}
\Delta_p^{AVu} \, &= \, \Delta_n^{AVd} \, &=\, 0.842, \nn \\
\Delta_p^{AVd} \, &= \, \Delta_n^{AVu} \, &=\, -0.427,  \label{axialff} \\
\Delta_p^{AVs} \, &= \, \Delta_n^{AVs} \, &=\, -0.085, \nn
\end{eqnarray}
while we use another set of coefficients for the tensor interaction:
\begin{eqnarray}
\Delta_p^{Tu} \, &= \, \Delta_n^{Td} \, &= 0.84, \nn \\
\Delta_p^{Td} \, &= \, \Delta_n^{Tu} \, &=-0.23, \\
\label{tensorff}
\Delta_p^{Ts} \, &= \, \Delta_n^{Ts} \, &=-0.046.  \nn 
\end{eqnarray}
The nucleon form factors for the axial-vector current, $\Delta_p^{AVu}$ and $\Delta_p^{AVd}$, have been provided by HERMES and COMPASS experiments \cite{Airapetian:2006vy,Ageev:2007du} while tensor coefficients come from lattice calculations \cite{Aoki:1996pi}.

The coefficients $\alpha_q$ can be extracted from the full matrix elements for $\chi q$ scattering via Fiertz transformations. While this is relatively straightforward to do analytically, numerical implementations of Fiertz transformations are non-trivial. In the following section, we discuss the numerical method \maddm{} uses to compute the low energy coefficients $\alpha_q$ from the full matrix elements in the $\chi q$ scattering.

\subsection{Projection Operator Method for Extraction of Low Energy Coefficients}

The procedure for computing the low energy coefficients for DM-nucleon scattering in \maddm{} is similar to the procedure implemented in {{MicrOMEGAs}} \cite{Belanger:2008sj}
\footnote{For improved treatments on QCD effect in dark matter scattering off nucleon, see Ref. \cite{Drees:1993bu} and Ref. \cite{Hisano:2015bma} for recent development.}. 
At the quark level, the input quark-DM interacting Lagrangian at low energy can be re-written as a set of effective operators as follows:
  \beq
    {\mathcal{L}}_{Q^2=0} = \sum_{i} \alpha_{i} \, {\cal{O}}_{q}^i  \, ,
    \label{projEff}
  \eeq
where the operators ${\cal{O}}_{q}$ standing for $\chi q$ scattering are defined in Table \ref{EffTab}, and the index $i$ runs over all the relevant contributions. The list of operators can be separated in odd and even operators under the interchange of quarks and anti-quarks as follows:
  \begin{align}
    {\mathcal{L}}_{Q^2=0} &= \left({\mathcal{L}}_{SI}^e + {\mathcal{L}}_{SI}^o \right) + \left({\mathcal{L}}_{SD}^e + {\mathcal{L}}_{SD}^o \right) \nn \\
                        &= \sum_{q,s} \alpha_{q,s}^{SI} {\mathcal{O}}_{q,s}^{SI} + \sum_{q,s} \alpha_{q,s}^{SD} {\mathcal{O}}_{q,s}^{SD},
  \end{align}
where $q = u,d,s,c,b,t$ and $s$ stands for even ($e$) or odd ($o$) operator. The Fiertz transformation which gives $\alpha_{q,s}$ coefficients projects the matrix element for $\chi q \rightarrow \chi q$ onto the set of effective operators. Numerically, this is equivalent to taking the interference term of the matrix element for $\chi q \rightarrow \chi q$ scattering with the even effective operator matrix element for the same process. The projection over the effective operators will select either SI or SD contributions since Eq.~\eqref{projEff} is written in an orthogonal basis. 

The even and odd coefficients can be further separated by considering scattering off of both quarks and anti-quarks. The sum and difference of the even and odd coefficients can be written as:
  \begin{eqnarray}
    \alpha_{q,e} + \alpha_{q,o} = \frac{\left| {\mathcal{M}}^{q*} \cdot {\mathcal{M}}_{\rm eff}^{q,e} \right|}{\left|{\mathcal{M}}_{\rm eff}^{q,e} \right|^2},  \\
    \alpha_{q,e} - \alpha_{q,o} = \frac{\left| {\mathcal{M}}^{\bar{q}*} \cdot {\mathcal{M}}_{\rm eff}^{\bar{q},e} \right|}{\left|{\mathcal{M}}_{\rm eff}^{\bar{q},e} \right|^2}, 
    \label{eq:alphas}
  \end{eqnarray}
where the $|{\cal{M}}_{\rm eff}^{q,e}|^2$ in the denominator takes into account the fact that the effective operators are not properly normalized. In Eq. \eqref{eq:alphas} we used the property that the odd coefficient changes sign under the interchange of $q \rightarrow \bar{q}$.  Note that the procedure of projection operators for extracting $\alpha_{q,s}$ is completely generic although the list of effective operators is non-exhaustive. Indeed, operators such as $\bar{\psi} \gamma_5 \psi$ are neglected since they are suppressed in the zero momentum limit.

\begin{table}[t]
  \centering
  \begin{tabular}{|c|c|c|c|}
  \hline
   & DM spin & Even & Odd \\
  \hline
  \multirow{4}{*}{SI} & & scalar current & vector current \\
  & $0$ & $2M_{\chi} S S^{*} \bar{\psi}_q \psi_q $ & $i \left( \partial_{\mu} S \, S^{*} - S \partial_{\mu} S^{*} \right) \bar{\psi}_q \gamma^{\mu} \psi_q$ \\
  & $1/2$ & $\bar{\psi}_{\chi} \psi_{\chi} \bar{\psi}_q \psi_q$ & $\bar{\psi}_{\chi} \gamma_{\mu} \psi_{\chi} \bar{\psi}_q \gamma^{\mu} \psi_q$ \\
  & $1$ & $2M_{\chi} A_{\chi \mu}^{*} A_{\chi}^{\mu} \bar{\psi}_q \psi_q$ & $i \left( A_{\chi}^{* \alpha} \partial_{\mu} A_{\chi \alpha} - A_{\chi}^{\alpha} \partial_{\mu} A_{\chi \alpha}^{*} \right) \bar{\psi}_q \gamma_{\mu} \psi_q $  \\
  \hline
  \hline
  \multirow{4}{*}{SD} & & axial-vector current & tensor current \\
  & $1/2$ & $\bar{\psi}_{\chi} \gamma^{\mu} \gamma^5 \psi_{\chi} \bar{\psi}_{q} \gamma_{\mu} \gamma_5 \psi_{q}$ & $- \frac12 \bar{\psi}_{\chi} \sigma_{\mu \nu} \psi_{\chi} \bar{\psi}_{q} \sigma^{\mu \nu} \psi_{q}$ \\
  & $1$ & $ \quad \sqrt{6} \left( \partial_{\alpha} A_{\chi \beta}^{*} A_{\chi \nu} - A_{\chi \beta}^* \partial_{\alpha} A_{\chi \nu} \right) \epsilon^{\alpha \beta \nu \mu} \bar{\psi}_q \gamma_5 \gamma_{\mu} \psi_q \quad $ & $ \quad i \frac{\sqrt{3}}{2} \left( A_{\chi \mu} A_{\chi \nu}^* - A_{\chi \mu}^* A_{\chi \nu} \right) \bar{\psi}_q \sigma^{\mu \nu} \bar{\psi}_q \quad $  \\
  \hline
  \end{tabular}
  \caption{List of effective operators taken from Ref. \cite{Belanger:2008sj}, implemented into \maddm{}. $S$, $\psi_{\chi}$ and $A_{\chi}$ correspond to scalar, fermion and vector DM fields, respectively.  }
  \label{EffTab}
\end{table}

\subsection{Differential Nucleus Recoil Rate}\label{sec:DDM}
The direct detection module of \maddm{} allows the user to calculate quantities relevant for DM-nucleus scattering beyond the spin-independent and spin-dependent cross section. One is, for instance, often interested in studies of nucleus recoil rates, where traditionally the quantity of interest has been $dR/dE, $ the differential nuclear recoil rate as a function of recoil energy. However, valuable information is contained in the angular distribution of the recoil rate $dR/d\Omega$. And despite the fact that no currently operating experiment has the ability to efficiently measure directionality of the nuclear recoil with high statistics, there are few ongoing efforts in this direction as well as concrete proposals for future experiments \cite{Grothaus:2014hja,Santos:2013oua,Daw:2013waa,Sciolla:2008ku,Nygren:2013nda,Mohlabeng:2015efa}.  Hence, we deemed it important to include the ability to calculate the angular rates into the \maddm{} code.

One reason why angular information could be important is that low energy recoil events recorded by direct detection experiments lie well within the range dominated by cosmogenic backgrounds and those resulting from radioactivity.
A strong signature of dark matter may come from directional detection, which exhibits a very large forward-backward asymmetry ($A_{FB}$) in the angular distribution of nuclear recoil events.  A large $A_{FB}$ would be difficult to mimic with any known backgrounds as one expects the angular distribution of such processes to be roughly uniform 

The fact that background processes are (typically) spatially isotropic could be used to overcome another impeding problem. As the direct detection bounds on the DM-nucleon cross-sections get lower, neutrino backgrounds will present an obstacle \cite{Billard:2013qya} and directional detection could provide a way to circumvent the so called ``neutrino floor'' \cite{Davis:2014ama}.  

In case a signal is ever observed, the study of directionality in DM-nucleus recoils will be of foremost interest. The angular distribution of the recoil rates could provide useful information about the astrophysical aspects of the DM galactic halo profile~\cite{Copi:1999pw, Copi:2000tv}, as well as information on the anisotropy in the DM halo velocity distribution and prospects for more accurate measurements of the DM mass and interaction cross-section~\cite{Mohlabeng:2015efa}.

\begin{figure}[t]
\includegraphics[scale=0.85]{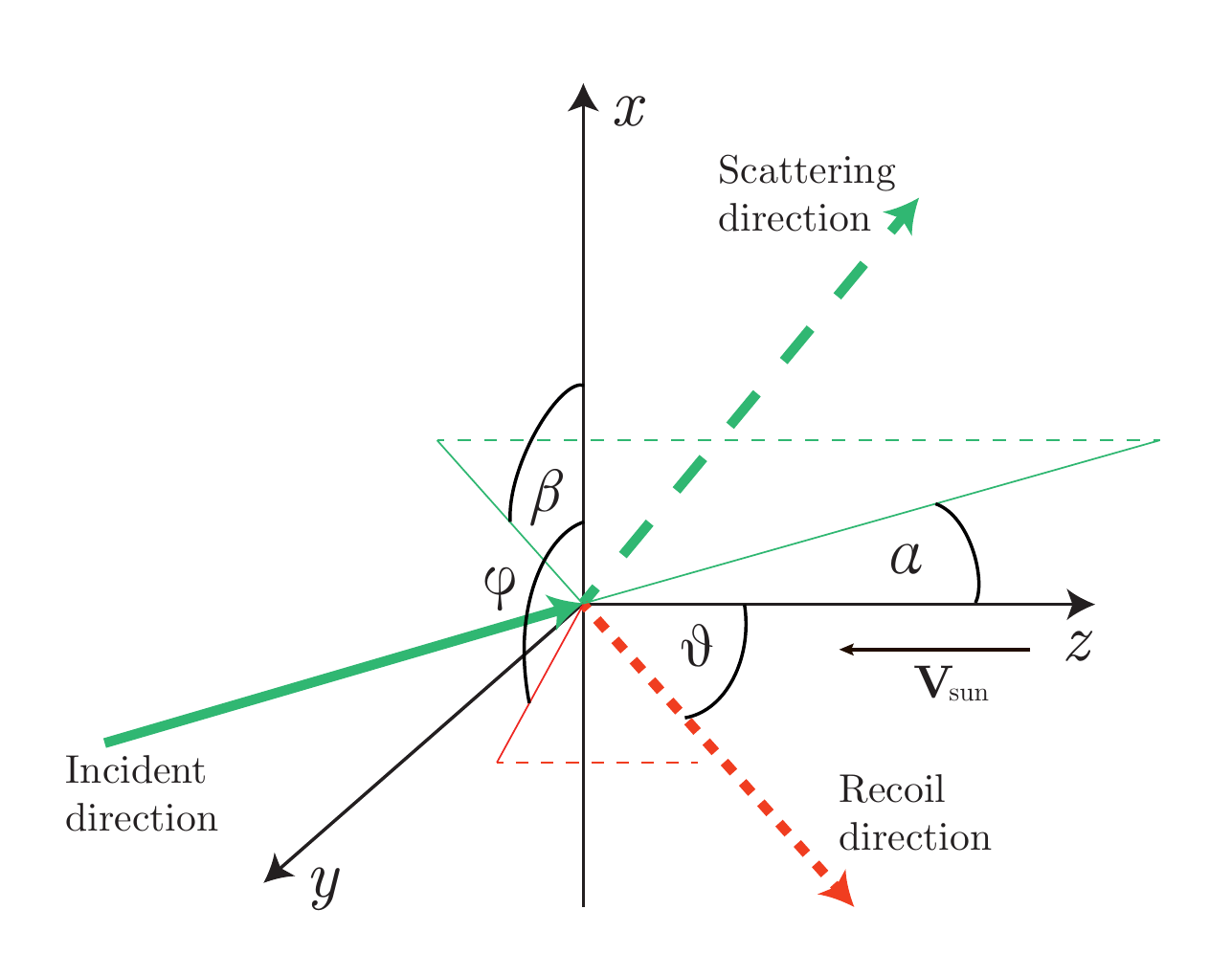}
\caption{Geometry of DM scattering off a target nucleus in the detector. The DM is incident at an angle $(\alpha, \beta)$ relative to the $\bf{z}$ axis. The nucleus recoils in the $(\theta, \phi)$ direction. Note that $\beta$ and $\phi$ angles lie in the $x,y$ plane. The thin and dashed thin lines serve to illustrate the projection of the incoming and recoil vectors onto the $z$ axis and the $xy$ plane.}
\label{DMd}
\end{figure}
In this section we present a detailed overview of the theoretical background of differential recoil rates for DM detection. We begin by considering our solar system moving through the Galactic ``DM-wind" in the direction of the Cygnus $X-2$ constellation as illustrated in Fig. \ref{DMd}. A DM particle of mass $m_{\chi}$, incident at velocity $\mbf{v} = v ( \sin \alpha \cos \beta \, \mbf{ \hat {x} } + \sin \alpha \sin \beta \, \mbf{\hat{y}} + \cos \alpha \, \mbf{ \hat{z} }) $ in the detector, elastically scatters off a target nucleus of mass $m_{N}$, stationary in the detector, and causes it to recoil with velocity $ \mbf{u} = u(\sin \theta \cos \phi \, \mbf{ \hat{x} } + \sin \theta \sin \phi \, \mbf{ \hat{y} } + \cos \theta \, \mbf{ \hat{z} }) $ and momentum $\mbf{q}$ in the direction $(\theta, \phi)$. The event rate per unit recoil energy and unit recoil solid angle is
\begin{eqnarray}
  \frac{d^{2}R}{dE_R d\Omega_{(\theta, \, \phi)}} = \frac{ 2 \, \rho_{0} }{ m_{\chi}} \int \frac{d\sigma}{dq^{2} d\Omega_{(\theta, \, \phi )}} v f({\bf v}) d^{3}v,
  \label{doublerate}
\end{eqnarray}
where $\rho_{0}$ is the dark matter density in our local galaxy and $f({\bf v})$ is the velocity distribution of the DM in the galactic halo.  The differential cross-section
\begin{equation}
 \frac{d\sigma}{dq^{2} d\Omega_{(\theta,\phi)}} = \frac{\sigma_{\chi N}}{8\pi \mu^{2} v} F^{2}(q) \delta \big (v \cos\theta - \frac{q}{2\mu} \big ),
\end{equation}
is obtained from the two body scattering phase space distribution \cite{Gondolo:2002np}. Here, $\mu = m_{\chi} m_{N}/(m_{\chi} + m_{N})$ is the reduced mass of the DM-nucleus system, $q =\sqrt{2 \,m_{N} \, E_{R}}$ is the recoil momentum with $E_{R}$ the recoil energy, angle $\theta$ is the recoil angle which determines the direction between the recoiling nucleus and the initial DM trajectory, while $\sigma_{\chi N}$ is the DM-nucleus scattering cross-section, weighted by the nuclear elastic scattering form factor $F(q)$. Assuming the nucleus is a sphere with uniform density, the form-factor is the Fourier transform of the nuclear density and includes all the relevant nuclear effects. For a detailed discussion on $F(q)$ see Section \ref{secFormFactor}. The DM-nucleus cross-section is summed over contributions from 
the spin-independent ($\sigma_{SI}$) and spin-dependent ($\sigma_{SD}$) cross-sections respectively, as defined in Eq.~\eqref{SIcx} and Eq.~\eqref{SDcx} respectively. 
Eq.~\eqref{doublerate} becomes 
\begin{eqnarray}
  \frac{d^{2}R}{dE d\Omega_{(\theta,\phi)}} = \frac{N_{0} \, \rho_{0} \, \sigma_{\chi N}}{\pi \, A \, r \, m_{\chi}^{2}} F^{2}(q) \int \delta(v \cos \theta - \frac{q}{2 \mu}) f({\bf v}) d^{3}v, 
\label{doublerate2}
\end{eqnarray}
where $r = 4 m_{N} m_{\chi}/(m_{N} + m_{\chi})^{2}$,  $N_{0} = 6.022\times10^{26} \kg^{-1}$ is Avogadro's number, $m_{N} = 0.932$ A is the target mass and A the atomic mass number (the factor of 0.932 is the value of atomic mass units (AMU) in GeV). This form of the double differential recoil rate is particularly useful as the quantity
\begin{equation}
\hat{f}(v_{q}, \mbf{\hat{q}}) = \int \delta(\mbf{v}\cdot\mbf{\hat{q}} - v_{q}) f(\mbf{v}) d^{3} v, 
\label{radon}
\end{equation}
is easily identified as the three dimensional Radon transform\footnote{For more information about the Radon transforms as applied to directional detection, we refer the reader to Ref.~\cite{Gondolo:2002np}.}, where  $ \mbf{\hat{q}}$ is the recoil momentum direction, $\mbf{v}$ is the velocity of the DM particle and $v_q= \frac{q}{2 \mu}$ is the minimum velocity required for the DM to impart a recoil momentum 
$\mbf{q}$ on the nucleus. $\hat{f}(v_{q}, \mbf{\hat{q}})$ represents the velocity distribution for a detector stationary in the galactic frame. We consider an observer moving through the Galaxy with velocity ${\bf V_{\rm lab}}$ in the galactic frame. ${\bf V_{\rm lab}}$ is related to the DM velocity in the lab frame (${\bf v_{\rm lab}}$) and the DM velocity in the galactic frame (${\bf v_{\rm gal}}$) by the Galilean transformation ${\bf v_{\rm lab} = \bf v_{\rm gal} - \bf V_{\rm lab}}$. The transformation takes into account the motion of a detector relative to the DM particles as seen in the lab frame. By the properties of the Radon transform for a pure translation $\bf V_{\rm lab}$, we obtain 
\begin{equation}
 \hat{f}_{\rm lab}(v_{q}, \mbf{\hat{q}}) =  \hat{f}_{\rm gal}(v_{q} + \mbf{V_{\rm lab}}.\mbf{\hat{q}}, \mbf{\hat{q}}).
 \label{radonlab}
\end{equation}
We assume a Maxwell-Boltzmann velocity distribution which is truncated at the DM escape velocity $v_{\esc}$ such that
\begin{equation}
  f_{{\rm MB}}(v) =  \frac{1}{k_{\esc} \pi^{3/2} v_{0}^{3}}  e^{-v^{2}/v_{0}^{2}}, 
  \label{maxwell}
\end{equation}
for $v < v_{\esc}$ and $ f_{\rm MB}(v) = 0$ otherwise. In Eq. (\ref{maxwell}), $v_0$ is the most probable speed of the DM in the halo with typical values of 220 $\sim$ 250 km/s and $k_{\esc}$ is a normalization factor which is obtained by integrating the velocity distribution in the galactic frame from 0 to $v_{\esc}$.
For ${\bf V_{\rm lab}} = {\bf v_{E}}$, where ${\bf v_{E}}$ is the velocity of the Earth with respect to the galactic frame, Eq. (\ref{radonlab})  becomes
\begin{equation}
  \hat{f}_{\rm MB}(v_{q} + \mbf{v_{E}}.\mbf{\hat{q}}, \mbf{\hat{q}})=  \frac{1}{k_{\esc} \sqrt{\pi} v_{0}} \left[{\rm exp}\left(-\frac{ (v_{q} + \mbf{v_{E}}.\mbf{\hat{q}})^{2}}{v_{0}^{2}}\right) - {\rm exp}\left(-\frac{v_{\esc}^{2}}{v_{0}^{2}}\right) \right].
  \label{maxwellradonmove}
\end{equation}
Combining Eq.~\eqref{doublerate2} and Eq.~\eqref{maxwellradonmove} we obtain an expression for a double differential recoil rate in the laboratory frame:
\begin{eqnarray}
  \frac{d^{2}R}{dE_{R} d\cos \theta} = \frac{2 \, N_{0} \, \rho_{0} \, \sigma_{\chi N} }{\pi^{1/2} \, A \, r \, v_{0} \, M_{\chi}^{2} } \frac{F^{2}(E_{R})}{k_{\esc}} \, \left[ \exp\left(-\frac{(v_{E} \cos \theta - v_{\min})^{2}}{v_{0}^{2}}\right) -\exp\left(-\frac{v_{\esc}^{2}}{v_{0}^{2}}\right) \right], 
\label{doubleratefin}
\end{eqnarray}
 where $ \mbf{v_{E}} \cdot \mbf{\hat {q}} = - v_{E}\cos \theta$ and $v_{q} = v_{\rm min} = v_0 \sqrt{E_{R}/E_{0} r} $, with $E_{0} = 1/2 \, m_{\chi} v_{0}^{2}$ being the most probable kinetic energy of the DM. The velocity of the Earth $v_{E}$, with respect to the galactic frame is calculated in the appendix of Ref.~\cite{Lewin:1995rx} and includes the motion of the Earth with respect to the Sun, the proper motion of the Sun and the motion of the solar system with respect to the galactic center. Eq. (\ref{doubleratefin}) assumes azimuthal symmetry of the DM velocity profile as illustrated in Fig.~\ref{DMd}. 

\subsection{Nuclear Form Factors and Detector Resolution}
\label{secFormFactor}

In order to compute the recoil rates, we take into account both spin-dependent (axial vector) and spin-independent (scalar) DM-nucleon interactions. The contributions of different types of interactions show up in the DM-nucleus cross-section and in the form-factor.  For the spin-independent case, we consider the Helm form factor as described in Ref. \cite{Lewin:1995rx}:
\begin{eqnarray}
 F_{SI}(q) = \frac{3[{\rm sin}(qr_{n})-qr_{n}{\rm cos}(qr_{n})]}{(qr_{n})^{3}}e^{-(qs)^{2}/2}, 
 \label{formfac}
\end{eqnarray}
with  $\rm q = 6.92\times10^{-3} {\rm A}^{1/2} E_{R}^{1/2}$ ${\rm fm}^{-1}$ the recoil momentum,  $r_{n} =  \sqrt{c^{2} + 7/3 \pi^{2}a^{2}-5s^{2}}$ fm the effective radius of the nucleus,  $\rm c = (1.23 A^{1/3} - 0.60)$ fm, $ a = 0.52$ fm and $ s = 0.9$ fm. 
Eq. (\ref{formfac}) is in fact the Wood-Saxon form factor, with $ r_{n} = c^{2} - 5s^{2}$ and $ c = 1.2 {\rm A}^{1/3}$ fm. 
Furthermore for the spin-dependent case, we consider
\begin{equation}
 F_{\rm SD}^{2} (q)= \frac{S(q)}{S(0)},
 \label{formsd}
\end{equation}
where the recoil momentum dependent structure function is given by $ S(q) = a_{0}^{2} S_{00}(q) + a_{0} a_{1} S_{01}(q) + a_{1}^{2} S_{11}(q)$ and $ S(0)$ is the structure function at zero momentum transfer. For the coefficients 
\begin{eqnarray}
a_{0} = \frac{11}{18} (\Delta_{u} + \Delta_{d}) + \frac{5}{18} \Delta_{s}\,,  \\
a_{1} = \frac{11}{18} (\Delta_{u} + \Delta_{d}) + \frac{5}{18} \Delta_{s}\,, 
\end{eqnarray}
 we use the nucleon form factors, $\rm \Delta_{\it i}$ ($ i = u, d, s$), as defined in Eq. (\ref{axialff}).
 The nuclear spin structure functions $S_{ij}$ for different nuclei are approximated using fitting functions. For Xenon, Iodine and Sodium we use the functions from the Nijmegen II fitting as calculated in Ref. \cite{Ressell:1997kx}. For Germanium we use those calculated in Ref. \cite{Dimitrov:1994gc}, while for Fluorine and Silicon we use the fitting functions found in Ref. \cite{Bednyakov:2006ux}. All the information we use for both spin-independent and spin-dependent interactions are summarized in Table \ref{targetsi} and Table \ref{targetsd}\footnote{We were also not able to find the values in asterisk. For Neon, Argon and Sulphur, we set the orbital spin values to $\frac{1}{2}$ to avoid any infinities in the calculation of the spin-dependent cross-sections as shown in Eq. \eqref{SDcx}, while we set the default values of the magnetic moments for Argon and Sulphur to 0.}  respectively. For the nuclear structure functions that are currently not very well calculated (such as Carbon, Sulphur, Neon and Argon), we use the Gaussian distribution $ S_{ij}(q) = S_{ij}(0) \,\mathrm{\exp}\left [ - \frac{q^{2} R_{\rm A}^{2}}{4} \right ]$ as described in Ref. \cite{Belanger:2008sj} with $ R_{\rm A} = 1.7 {\rm A}^{1/3} - 0.28 - 0.78(\rm A^{1/3} - 3.8 + \sqrt{(\rm A^{1/3} - 3.8)^{2} + 0.2})$ for $\rm A \, \textless \,100$ and $ R_{\rm A} = 1.5 \rm A^{1/3} $ for $ \rm A \, \textgreater \,100$. 
 We note that the Gaussian approximation breaks down for large recoil energies \cite{Belanger:2008sj}.

\begin{table}[t!]
\centering
\begin{tabular}{|c|c|c|}
\hline
Target Material &  & List of Stable Isotopes $\&$ Abundances  \\ \hline
\hline
Xenon & $A$ & 123.9, 125.9, 127.9, 128.9, 129.9, 130.9, 131.9, 133.9, 135.9  \\
  Z =  54  & Abundance (\%) & 0.09, 0.09, 1.92, 25.44, 4.08, 21.18, 26.89, 10.44, 8.87  \\ \hline
Germanium & $A$ & 69.92, 71.92, 72.92, 73.92, 75.92  \\
  Z =  32  & Abundance (\%) & 20.84, 27.54, 7.73, 36.28, 7.61  \\ \hline
Silicon & $A$ & 27.98, 28.98, 29.97  \\
  Z =  14  & Abundance (\%) & 92.23, 4.68, 3.087  \\ \hline
Neon & $A$ & 19.99, 20.99, 21.99  \\
  Z =  10  & Abundance (\%) & 90.48, 0.27, 9.25  \\ \hline
Argon & $A$ & 37.962, 39.962  \\
  Z =  18  & Abundance (\%) & 0.0632, 99.6  \\ \hline
Sodium & $A$ & 22.9  \\
  Z =  11  & Abundance (\%) & 100  \\ \hline
Iodine & $A$ & 126.9  \\ 
  Z =  53  & Abundance (\%) & 100  \\ \hline
Carbon & $A$ & 12.0, 13.0  \\ 
  Z =  6  & Abundance (\%) & 98.89, 1.11  \\ \hline
Fluorine & $A$ & 18.998  \\ 
  Z =  9  & Abundance (\%) & 100  \\ \hline 
Sulphur & $A$ & 31.9, 32.97, 33.96, 35.96  \\ 
 Z =  16 & Abundance (\%) & 94.9, 0.76, 4.29, 0.02  \\ \hline
\end{tabular}
\caption{List of stable isotopes and their abundances for the different target materials used in the calculation of the spin-independent quantities. \label{targetsi}}
\end{table}

\begin{table}[t!]
\centering
\begin{tabular}{|c|c|c|c|c|}
\hline
{  }Target Material {  } & {  }$S_{ij}$  {  }& {     }$\langle S_{n} \rangle $ {     }& {     }$ \langle S_{p} \rangle$ {     }& {  }{  }J{ } {  } \\ \hline
\hline
Xenon & Ref. \cite{Ressell:1997kx} & -0.272 & -0.009 & $\frac{3}{2}$ \\ \hline
Germanium & Ref. \cite{Dimitrov:1994gc} & 0.378 & 0.030 &  $\frac{9}{2}$\\ \hline
Silicon & Ref. \cite{Bednyakov:2006ux} & 0.13 & -0.002 & $\frac{1}{2}$ \\ \hline
Neon & Refs. \cite{Belanger:2008sj, Bednyakov:2006ux} & 0.294 & 0.02 &  $\frac{1}{2}^\ast$ \\ \hline
Argon & Ref. \cite{Belanger:2008sj} & $0^\ast$ & $0^\ast$ & $\frac{1}{2}^\ast$ \\ \hline
Sodium & Ref. \cite{Ressell:1997kx} & 0.0199 & 0.2477 &  $\frac{3}{2}$ \\ \hline
Iodine & Ref. \cite{Ressell:1997kx} & 0.075 & 0.309 &  $\frac{5}{2}$\\ \hline
Carbon & Ref. \cite{Belanger:2008sj} & -0.172 & -0.009 & $\frac{1}{2}$  \\ \hline
Fluorine & Refs. \cite{Bednyakov:2006ux, Divari:2000dc} & -0.0087 & 0.4751 & $\frac{1}{2}$ \\ \hline
Sulphur & Ref. \cite{Belanger:2008sj} & $0^\ast$ & $0^\ast$ & $\frac{1}{2}^\ast$ \\ \hline
\end{tabular}
\caption{\label{targetsd} List of structure functions, moments and angular momenta for different materials. The fitting polynomials for the structure functions can be found in the references cited.
We were not able to find the magnetic moments for Argon and Sulphur thus we set them to be zero, thus MadDM does not currently compute the SD rates for them. We were also not able to find the orbital spin values for Argon, Neon and Sulphur, we set these to $\frac{1}{2}$ in the code at the moment.}
 
\end{table}

Finally we use Eq. \eqref{doubleratefin} to simulate the scattering of dark matter with the nucleus in a detector. 
We take into account the energy and angular resolutions of a generic detector as applied to the distribution of the scattering events. 
We assume a Gaussian resolution function to smear the distribution of recoil events in both energy and angle 
with standard deviations $\sigma_{E} = \lambda \sqrt{E}$ (the energy resolution) and $\sigma_{\theta}$ (constant angular resolution) as in Ref. \cite{Mohlabeng:2015efa}. Default resolutions are given by $\lambda =1$ and $\sigma_\theta = 30^\circ$ with an option for users to implement their own detector resolutions.


\section{How to Use MadDM v.2.0}\label{sec:code}
\subsection{Running \maddm} \label{running}

Addition of the direct detection code to the existing \maddm{}  package follows the syntax and philosophy of the previous \maddm{} version. The code is split up into the \verb|Python| and the \verb|FORTRAN| modules, whereby diagram generation and the structure of folders and files is handled by \verb|Python|, and the numerical calculations are performed by the \verb|FORTRAN| module. As in the previous version, the code does not require any pre-compilation. Simply placing the code within a MadGraph5\_aMC@NLO folder suffices. Note however, that \verb|Numpy| is now a pre-requisite for running the \maddm v.2.0 code. In order to use the plotting functionality, \verb|Matplotlib| is also required.

In the following we briefly describe how to use the \maddm{} code relevant for direct detection of dark matter, while we give detailed descriptions of the main functions/features of the \maddm{} v.2.0 package in the Appendix \footnote{For the reference on the relic density calculation see Ref. \cite{Backovic:2013dpa}.}. 

As in version 1.0, the main \maddm{} program can be executed via the \verb|maddm.py| script. The interface will prompt the user to enter the name of the UFO dark matter model (TAB auto-complete included), a name of the projects, as well as offer a choice of available computations (relic density, DM-nucleon cross sections and recoil rates). The \verb|Python| module of \maddm{} will then proceed to generate all the relevant scattering/annihilation diagrams and set up the directory structure of the user defined project in \verb|Projects/<projectname>| folder. 

The user can choose to enter the DM candidate manually, or to allow \maddm{} to determine the dark matter candidate automatically (the code assumes that the lightest particle with the PDG code greater than 25 and zero width is the DM candidate). The code also offers the user to edit the default parameter card, and pre-set the model parameters before it searches for the DM candidate.

Upon determining the DM candidate, \maddm{} will attempt to find any co-annihilation candidates. The user is prompted to enter the mass splitting within which the co-annihilation partners are deemed to contribute significantly to the calculation of relic density (the default is 10 \%). 

Finally, the code will prompt the user to edit the \verb|maddm_card.inc| file, which contains all relevant parameters for the numerical calculations. These include, but are not limited to precision parameters for the relic density calculation, nucleon form factors for direct detection, choices of target materials and detector parameters. 
Upon execution and output of the results, the code will offer the user the option to perform a parameter scan. 
Alternative to the user interface in \verb|maddm.py| the \maddm{} code can also be run within any user defined \verb|Python| script. We provide such an example in the \verb|example.py| file.

\subsection{UFO Model Conventions} \label{UFOconvention}
 
Now we proceed to discuss some important conventions on UFO files required by the \maddm v.2.0 code.
We implemented the projection operator formalism in \maddm{} v.2.0 via a recent feature of \madgraph{} which allows the user to merge two UFO model files. We provide UFO models with vertices of effective operators as a part of the \maddm{} code,  so that an interference term between the matrix elements of the user defined model, and the effective operators can be extracted. 
 The vertices are then automatically added to the DM model provided by the user during the execution of the code.
The merging procedure of \madgraph{} is sensitive to the UFO convention and the user should take special care to make sure that the DM model used in \maddm{} is compatible with the effective operator models embedded in \maddm{} v.2.0\footnote{The statement is relevant only for direct/directional detection. Relic density will proceed correctly without regard for the UFO version. }. We have generated our set of effective operators with {FeynRules 2.0}. We advise the user to  test the merging procedure in \madgraph{} before running \maddm{} in order to avoid UFO conflicts. Using this UFO convention, the DM particle should be defined as
\begin{verbatim}
    DM = Particle(pdg_code = pdg code,
                  ...
                  spin = 1,
                  mass = Param.DMMASS,
                  width = Param.WDM,
                  ...)
\end{verbatim}  
in case of a real scalar DM (for example). In this convention, \textbf{it is compulsory to assign a PDG code to the DM particle as well as a mass and a width}. Next, the mass and width block parameters have to match the following template:
\begin{verbatim}
DMMASS = Parameter(name = 'DMMASS',
                   ...
                   lhablock = 'MASS',
                   lhacode = [ pdg code ])

   WDM = Parameter(name = 'WDM',
                   ...
                   lhablock = 'DECAY',
                   lhacode = [ pdg code ])
\end{verbatim}        
where the \verb|lhablock| and \verb|lhacode| parameters should be exactly like the template when replacing the PDG code by the PDG code of the DM particle. Note also that we cannot use \verb|mdm| for the DM mass name due to conflict with internal {\maddm{}} parameters. For the same reason, if the user wants to define new coupling orders, the following ones should not be used : \verb|SIEFFS|, \verb|SIEFFF|, \verb|SIEFFV|, \verb|SDEFFF|, \verb|SDEFFV|.

Finally, the user should \textbf{make sure that direct detection diagrams have a QED order lower or equal to 2} since \maddm{} generates diagrams with the \verb|QED=2| flag.


\section{Validations} \label{sec:validations}

We performed validations of the \maddm{} code in a diverse class of benchmark models. In each case, we verified that the code reproduces accurately the results from existing literature or public codes such as MicrOMEGAs. 
 
\subsection{Validations of the DM-nucleon Cross Section Calculation}

\subsubsection{Simplified DM Models}
The simplest benchmark model we consider for validation is a scalar DM which communicates to the SM via the Higgs boson.
The model contains only one diagram contributing to direct detection {\ie} a t-channel Higgs exchange. The Higgs-DM coupling is included in the $\frac{\delta}{2} H^{\dagger}HS$ operator where $S$ is the DM scalar field. It implies that the model leads only to spin-independent cross section. 

\begin{figure}[tb]
 \centering
 \subfigure{\label{HPSIp}
 \includegraphics[scale=0.4]{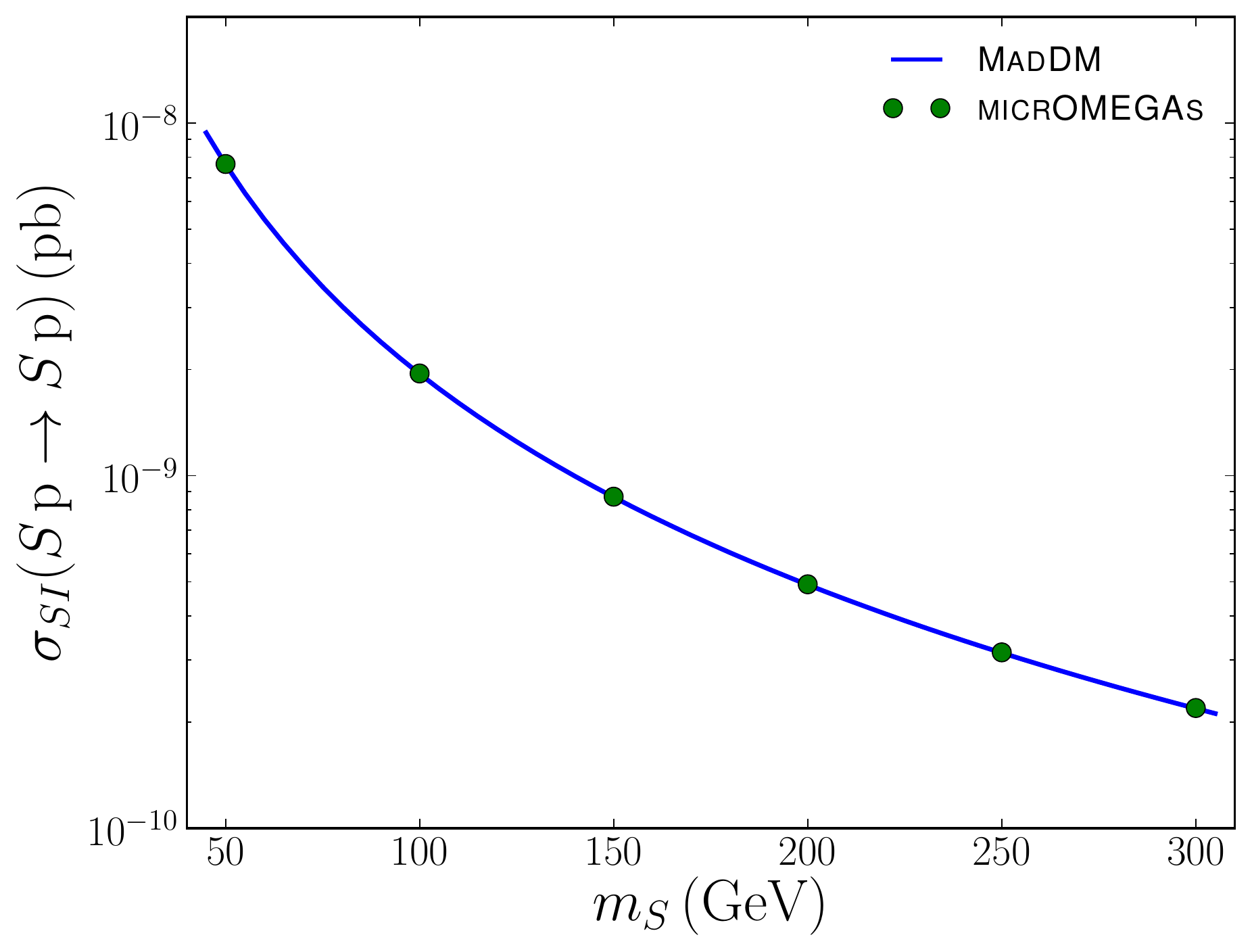}}
 \subfigure{\label{HPSIn}
 \includegraphics[scale=0.4]{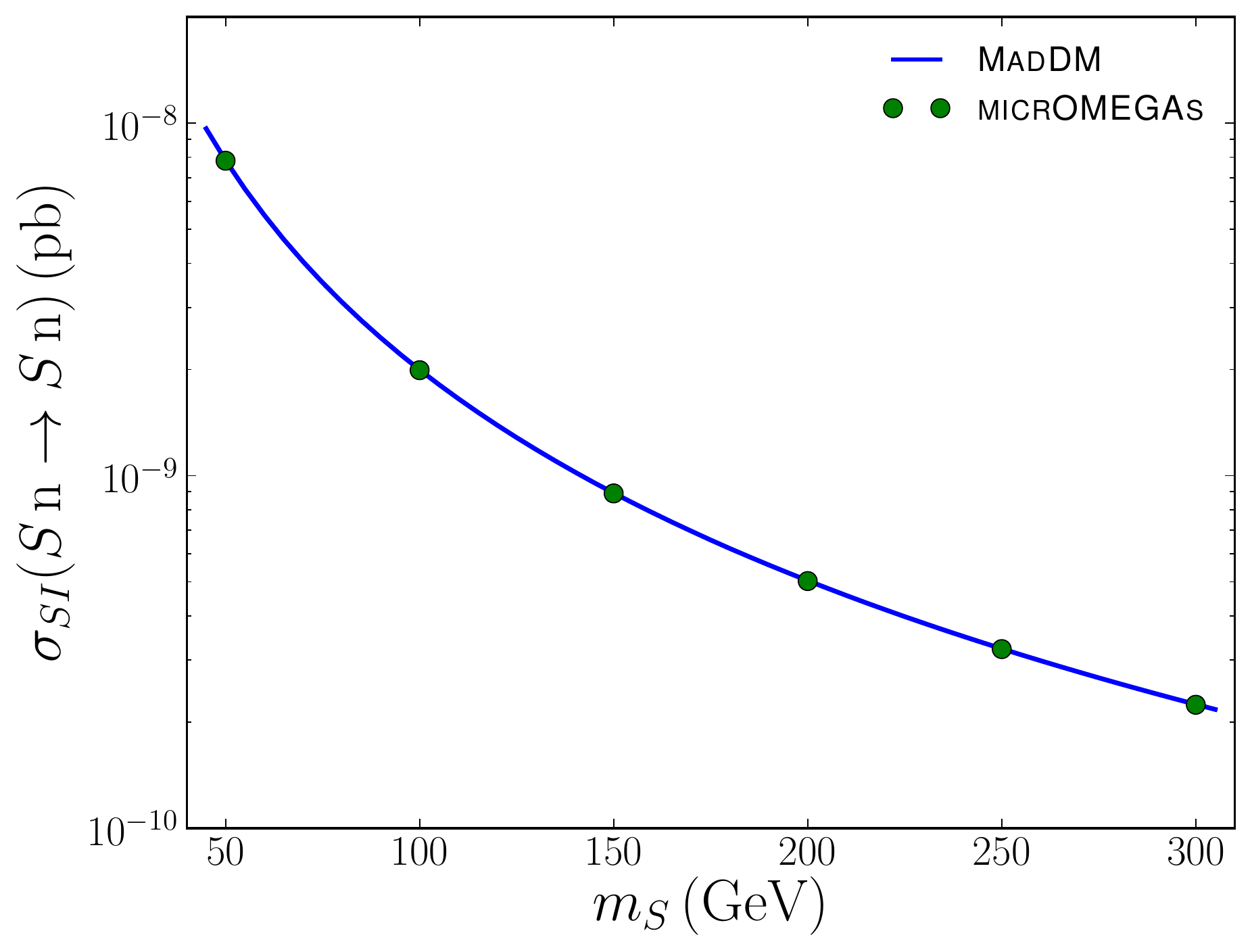}}\\
 \subfigure{\label{AVSDp}
 \includegraphics[scale=0.4]{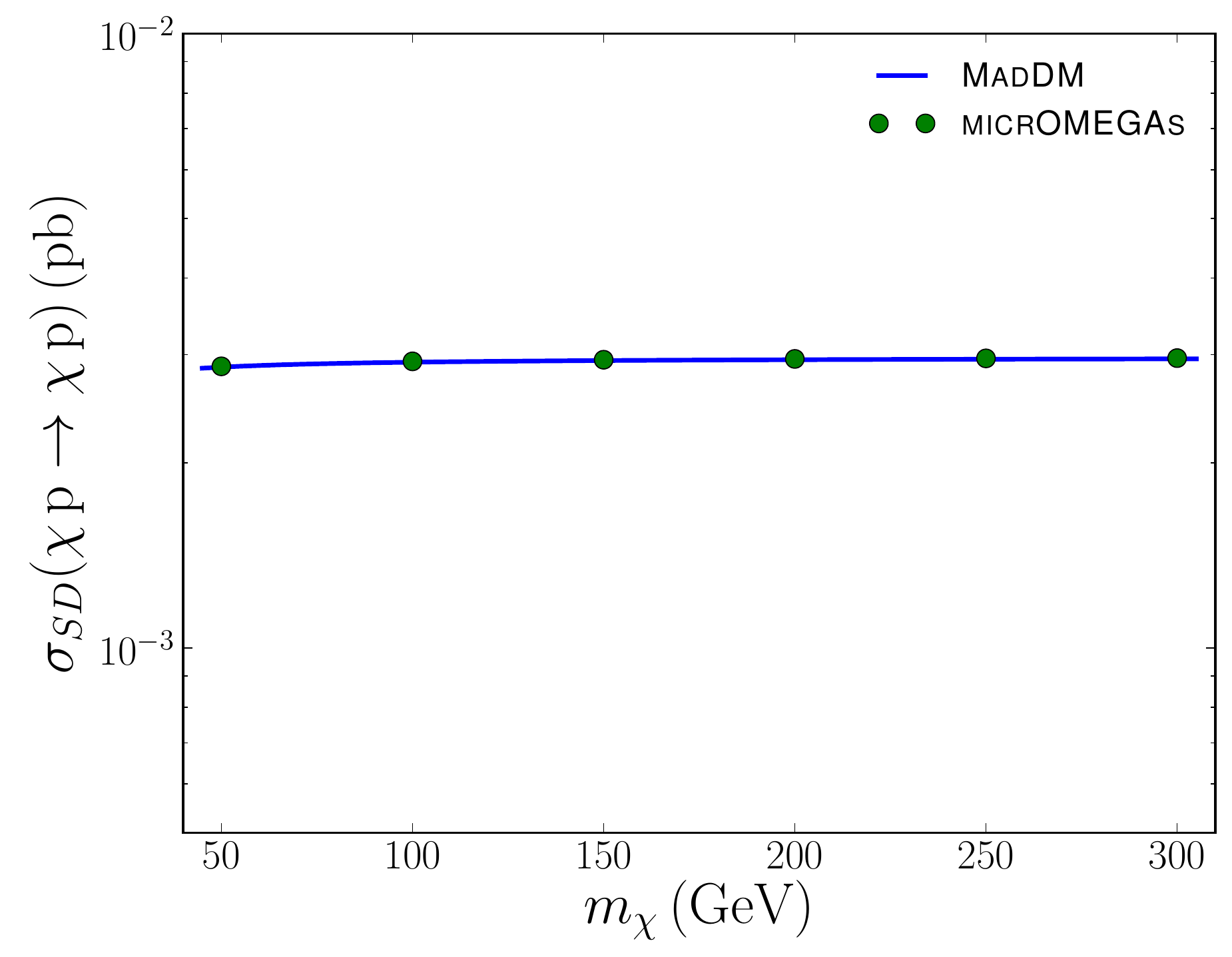}}
 \subfigure{\label{AVSDn}
 \includegraphics[scale=0.4]{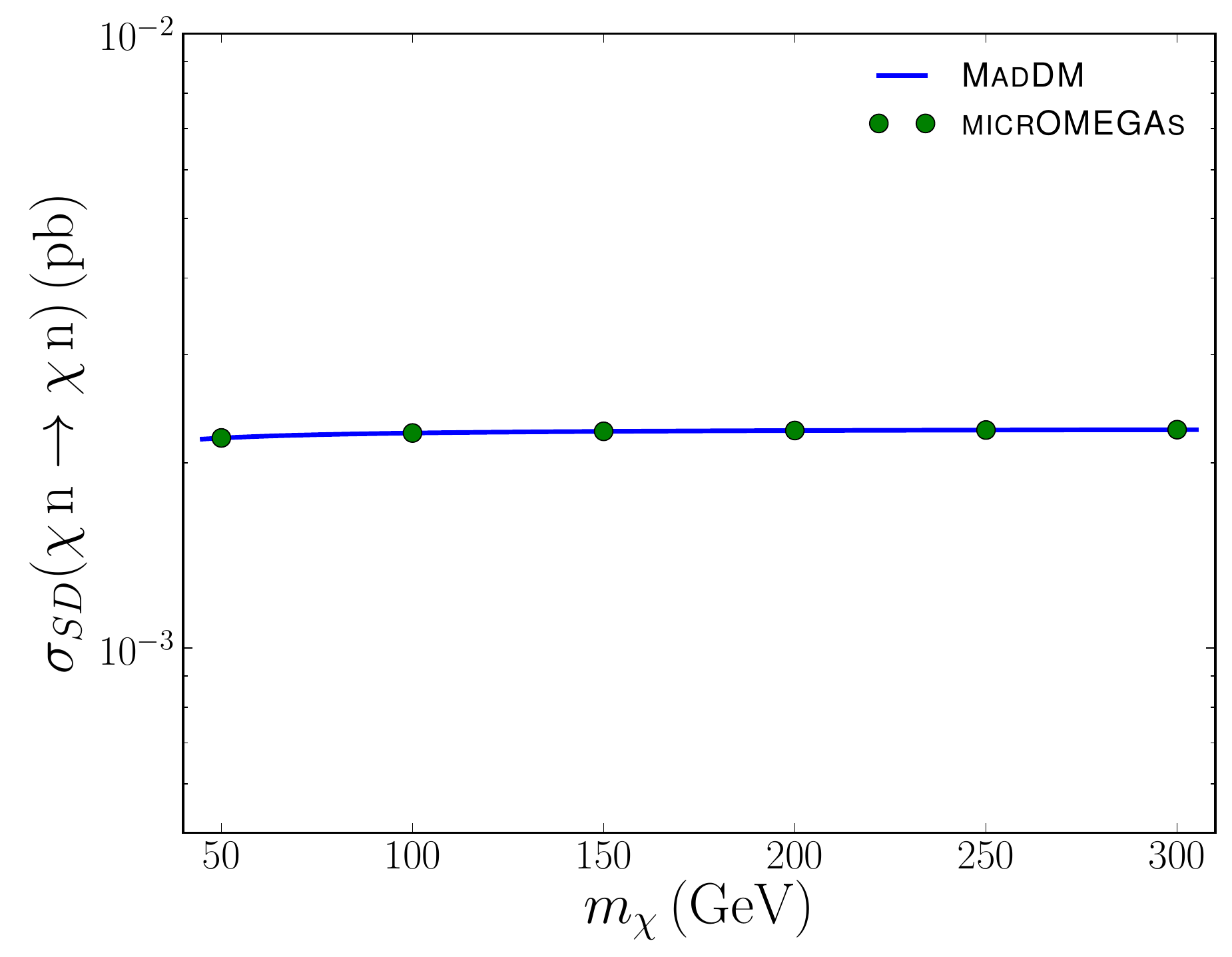}}
 \caption{Spin-independent elastic scattering cross section of scalar DM in the simplified SM model scenario with proton and neutron for $\delta = 0.1$ (top). Spin-dependent elastic scattering cross section of Dirac fermion DM with proton and neutron for $g_{\chi}^A = 0.1$ (bottom).}
 \label{SimpModels}
\end{figure}

For the purpose of validating calculations of the spin-dependent cross sections, we consider a similar simplified model,  with an axial-vector mediator which couples through fermionic DM and SM via the following Lagrangian:
\beq
{\mathcal{L}}_A = g_{\chi}^A \, \bar{\chi} \gamma^{\mu} \gamma_5 \chi V_{\mu} + g_q^A \, \bar{q} \gamma^{\mu} \gamma_5 q V_{\mu} \, , 
\eeq
where $\chi$ and $V_{\mu}$ are the DM and mediator fields, respectively. Fig. \ref{SimpModels} shows our results for proton/neutron DM cross section for the two simplified models against results obtained with MicrOMEGAs. In both cases, we find an excellent agreement between \maddm{} and MicrOMEGAs with the differences at $\lesssim 1 \%$ level.
We have also checked that the results from \maddm{} are consistent results from MicrOMEGAs in case of simplified models with different mediators (scalar, fermion and vector) and different DM candidate (scalar, fermion and vector), which are shown in Table \ref{EffTab}. 

\subsubsection{Minimal Universal Extra Dimensions}

After validating {\maddm{}} for a class of simplified models, we turn to more complex models. For this purpose, we chose to examine the results of \maddm{} for the Minimal Universal Extra Dimension model (MUED) against the results in existing literature. 

MUED is  the simplest model containing a Kaluza-Klein (KK) dark matter candidate among extra dimensions theories. 
The lightest KK particle, the KK-photon ($\gamma_1$) appears as a dark matter candidate in vanilla UED scenarios. At tree level, the inverse radius $R^{-1}$ of the extra-dimension corresponds roughly to the mass of the massive fields at level one. The mass splitting between KK-quark ($q_1$) and KK-photon:
\beq
\Delta = \frac{m_{q_1}-m_{\gamma_1}}{m_{\gamma_1}} \, ,
\eeq
plays an important role 
as a free parameter in direct detection. The details and phenomenology of MUED will not be discussed here since it has been studied extensively in the past \cite{Arrenberg:2008wy,Arrenberg:2013paa,Servant:2002hb,Cheng:2002ej} and we will proceed directly to the comparison of \maddm{} results on direct detection to the results in literature.

\begin{figure}[t]
 \centering
 \subfigure{\label{MUEDS}
 \includegraphics[scale=0.4]{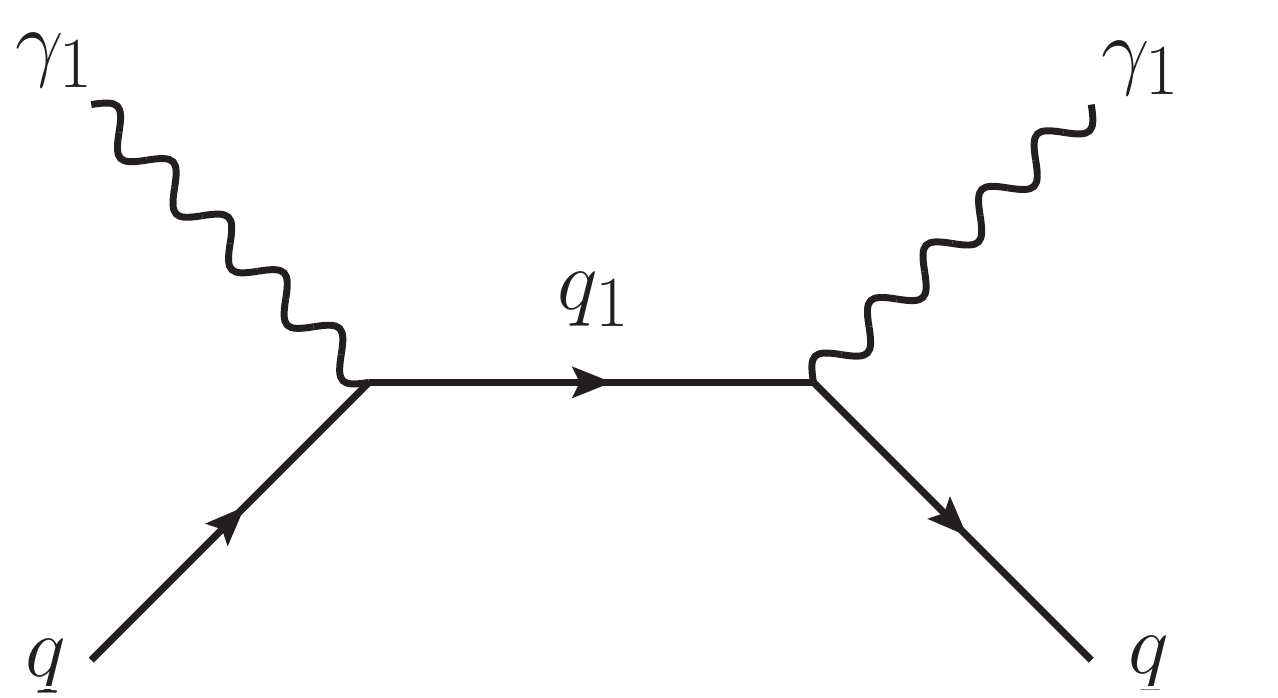}}
 \subfigure{\label{MUEDT}
 \includegraphics[scale=0.4]{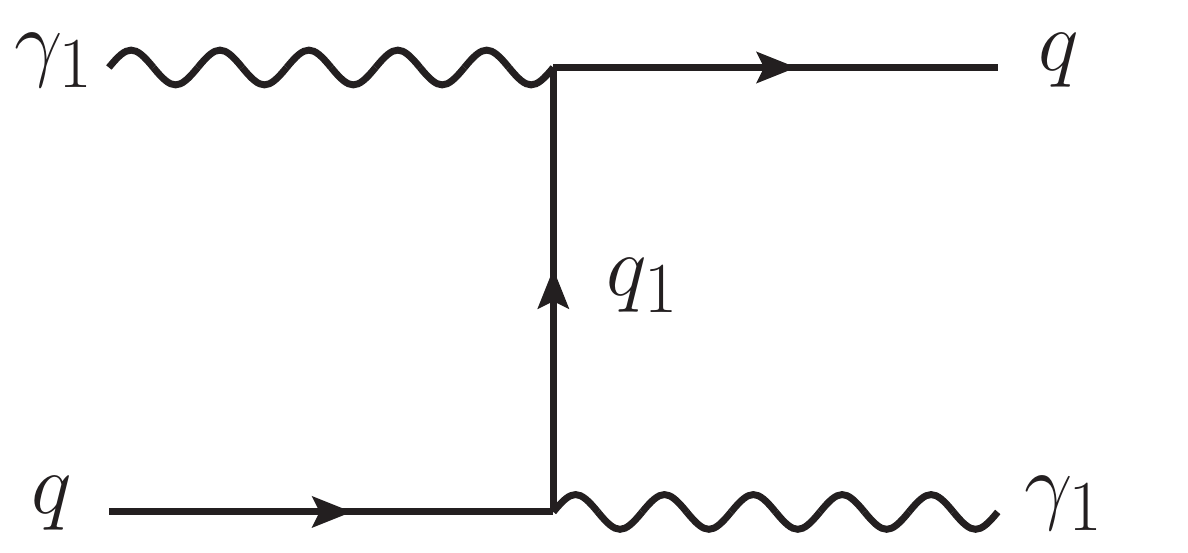}}
 \subfigure{\label{MUEDH}
 \includegraphics[scale=0.4]{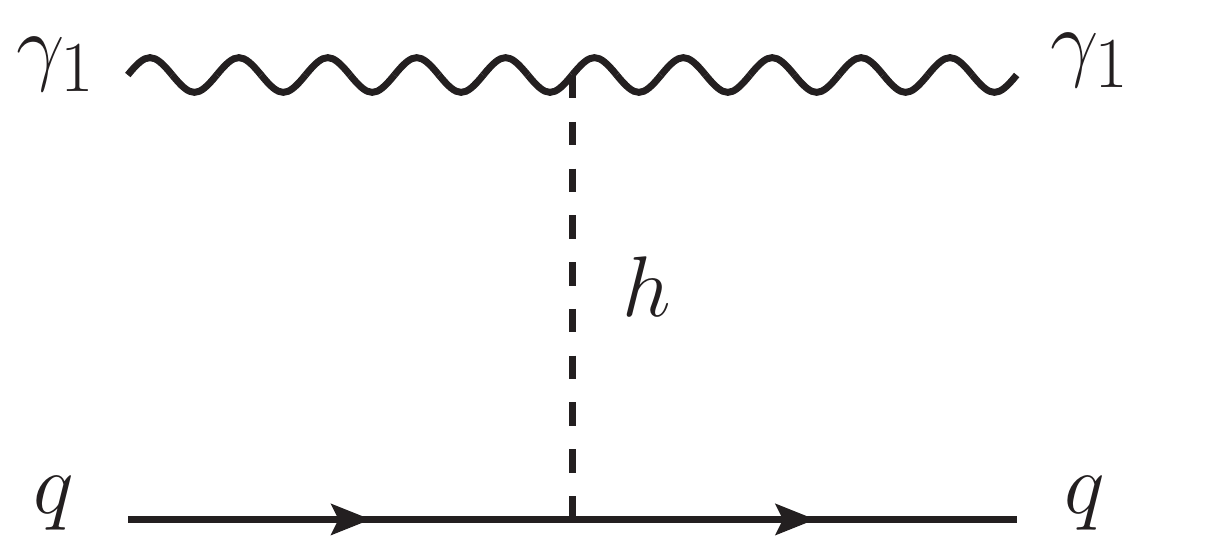}}
 \caption{Feynman diagrams for elastic scattering of KK photon with a quark.}
 \label{ued-diagrams}
\end{figure}

\begin{figure}[t]
\centering
\subfigure{\label{MUEDSI}
 \includegraphics[scale=0.4]{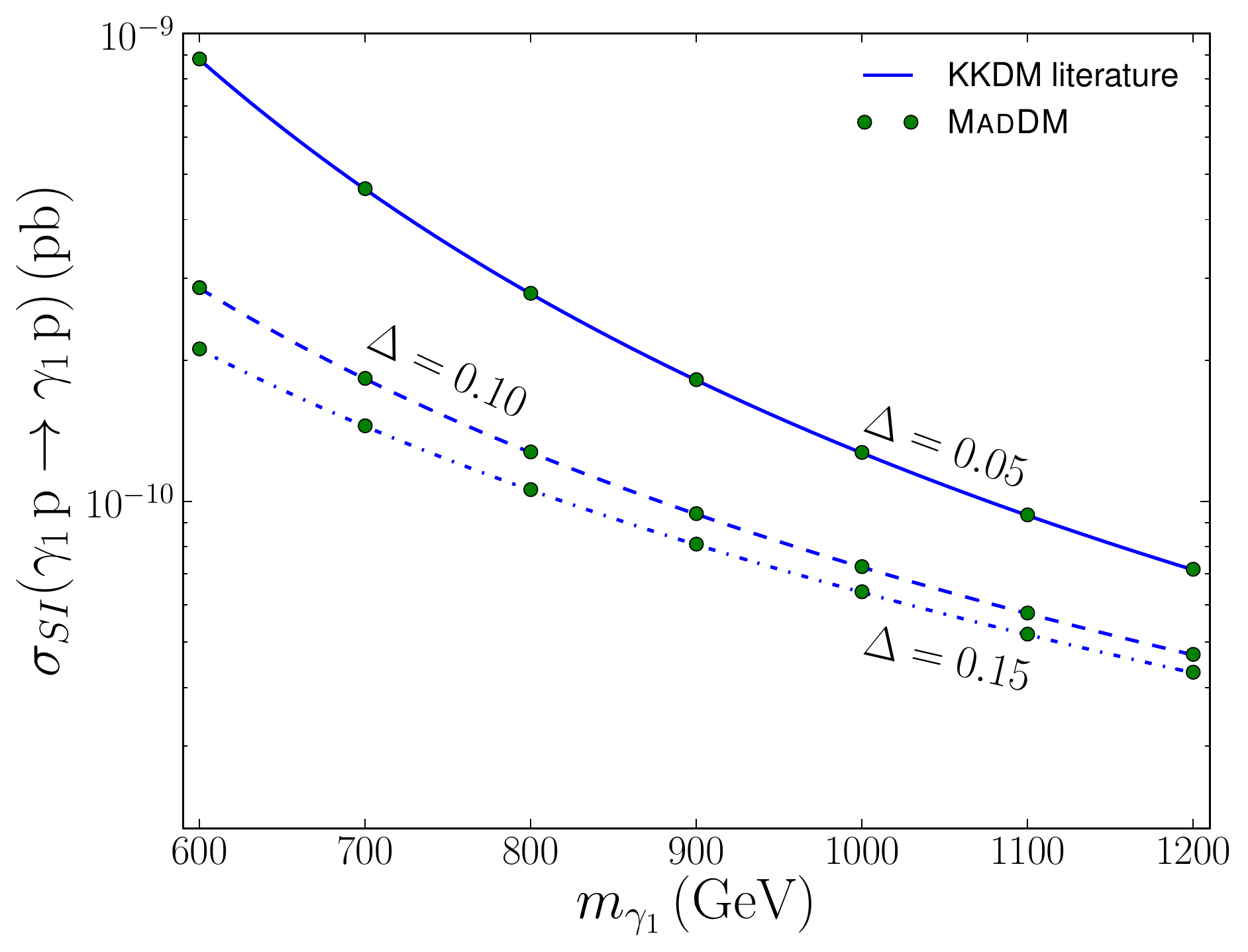}  \hspace{0.2cm}}
 \subfigure{\label{MUEDSD}
 \includegraphics[scale=0.4]{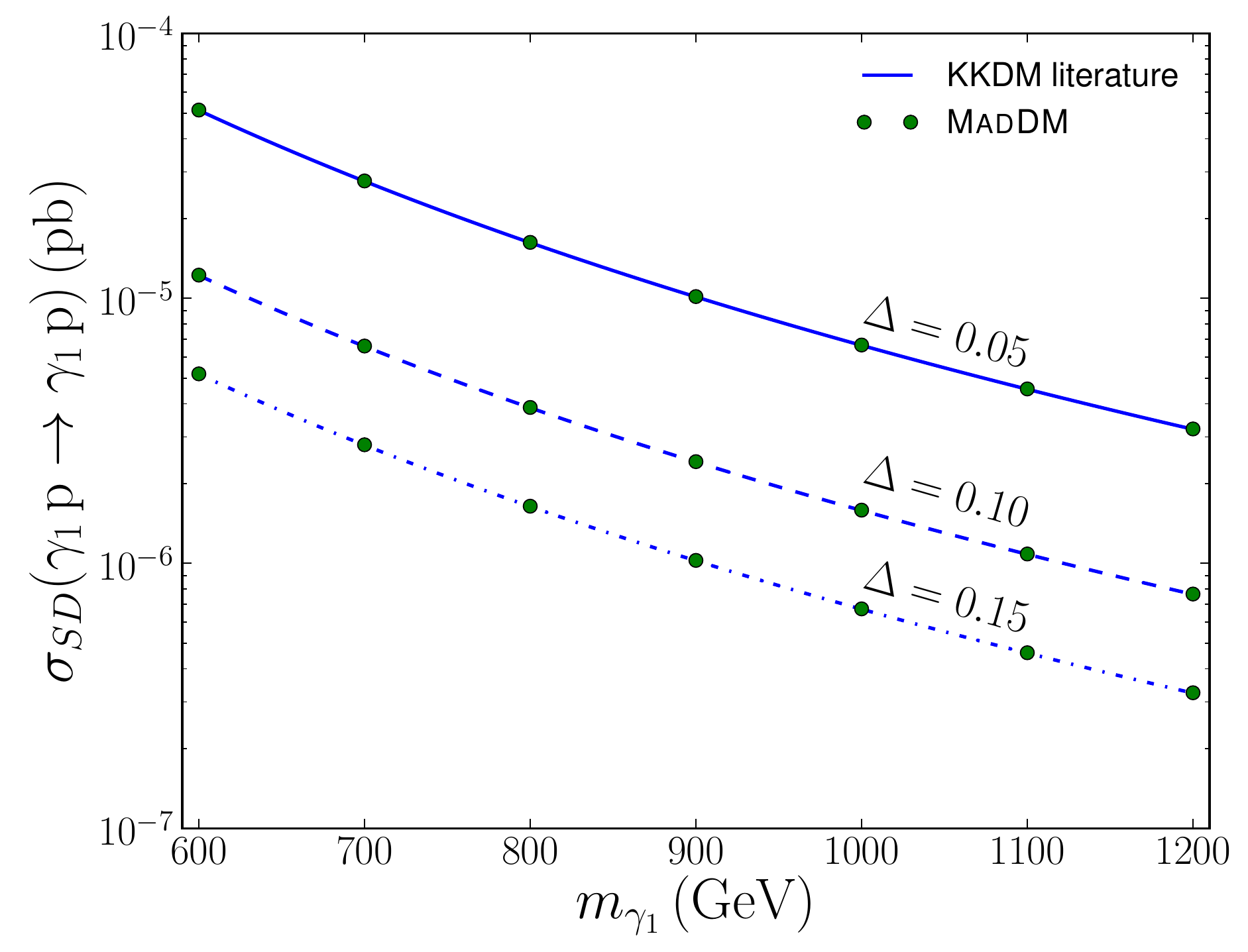} }
 \caption{Spin-independent (left panel) and spin-dependent (right panel) DM-proton cross section for MUED. The blue curves correspond to theoretical values coming from Kaluza-Klein dark matter literature \cite{Cheng:2002ej,Servant:2002hb}. The green points are the MadDM data. We show the results for three values of the $\Delta$ parameter.}
 \label{ued}
\end{figure}

The diagrams contributing to DM-nucleon cross section are displayed in Fig. \ref{ued-diagrams}. The Higgs-exchange diagram contributes to SI DM-nucleon cross section while the other two diagrams involving KK quarks contribute both for SI and SD DM-nucleon cross sections. 
We compared the \maddm{} results against those using the private code used in Refs. \cite{Arrenberg:2008wy,Cheng:2002ej}.
The results, shown in Fig. \ref{ued}, show perfect agreement between \maddm{} and the private code over a wide range of dark matter masses and for several values of $\Delta$. The DM-neutron cross sections are also in excellent agreement. \\

\subsubsection{Minimal Supersymmetric Standard Model}

Finally, we also validated \maddm{} for SPS1a benchmark point in Minimal Supersymmetric Standard Model (MSSM) \cite{Allanach:2002nj}.
MSSM neutralino dark matter is similar to KK photon dark matter in that the processes relevant for direct detection typically proceed via squark exchange in the $s$- and $u$-channels and Higgs exchange in the $t$-channel 
(see Fig. \ref{ued-diagrams}). In Table \ref{MSSMtable}, we compare {\maddm{}} with {{MicrOMEGAs}} for DM-nucleon cross-section for several SPS points. The ratio of the two tools cross-sections indicates that {\maddm{}} agrees with {{MicrOMEGAs}} to $\sim 1 \%$ level, with the exception of the SPS5 point. We have checked that the statement is also true if we vary the neutralino mass around the SPS1a point ($m_{\tilde\chi_1^0}=96.68 \, {\rm GeV}$). 

We note that it is challenging to compare other MSSM parameter points between MicrOMEGAs and {\maddm{}}. The default implementation of the MSSM model in MicrOMEGAs contains numerous model specific optimizations and improvements which we are not included in {\maddm{}}. Hence, the only way we could compare the two codes in an ``apples to apples'' fashion in the context of MSSM was to use the MSSM implementation provided on the FeynRules website. Despite numerous technical difficulties, we were able to generate MSSM models for several specific SPS points, by first pre-processing the SLHA cards from SUSYHIT to fit the format required by FeynRules and then using those parameter sets to output both UFO and CalcHEP models.  For the purpose of comparison, we manually turned off running of the QCD coupling in MicrOMEGAs by setting the variable \verb|qcdNLO=1| in \verb|directDet.c| and turned off the contributions from higher twist operators by setting \verb|Twist2On=0|, as these features are not built into {\maddm{}}.

\begin{table}[!]
  \centering
  \def\arraystretch{0.9}
\begin{tabular}{|c||c|c|c|}
\hline
$\sigma$ type & MadDM [pb] & micrOMEGAs [pb] & \% difference \\
\hline
\multicolumn{4}{c} {SPS1a}\\ \hline \hline

SI (proton) &2.16E-10  & 2.18E-10  &  0.9 \\
SI (neutron) &2.15E-10   &  2.17E-10 & 0.5 \\
SD(proton) &  6.53E-6 &  6.56E-6  & 0.5 \\
SD (neutron) &  8.79E-6 & 8.79E-6  & 0.0 \\
\hline
\hline
\multicolumn{4}{c} {SPS1b}\\ \hline
\hline
SI (proton) & 3.87E-11      & 3.89E-11  & 0.5  \\
SI (neutron) &3.85E-11    &  3.87E-11 & 0.4 \\
SD(proton) &  1.20E-6 &  1.21E-6  & 0.5 \\
SD (neutron) &   1.47E-6  & 1.47E-6  & 0.0 \\
\hline\hline
\multicolumn{4}{c} {SPS2}\\ \hline
\hline
SI (proton) & 7.35E-11 & 7.39E-11 & 0.5 \\
SI (neutron) & 7.35E-11 & 7.39E-11 & 0.5 \\
SD(proton) & 9.86E-6& 9.89E-6 & 0.3 \\
SD (neutron) & 8.05E-6 & 8.05E-6 & 0.0 \\
\hline\hline
\multicolumn{4}{c} {SPS3}\\ \hline
\hline
SI (proton)    &    5.55E-11   &  6.02E-11  & 0.4 \\
SI (neutron)  &   5.53E-11    &   5.98E-11 & 0.1 \\
SD(proton)    &    1.63E-6   &  1.64E-6 &  0.6\\
SD (neutron) &  2.08E-6     &  2.08E-6  & 0.0 \\
\hline\hline
\multicolumn{4}{c} {SPS4}\\ \hline
\hline
SI (proton)    &  2.04E-10      &  2.07E-10  & 1.3 \\
SI (neutron)  &   2.03E-10     &  2.05E-10 &  1.0\\
SD(proton)    &  7.56E-6      &  7.58E-6  & 0.3 \\
SD (neutron) &   7.82E-6     &   7.82E-6 & 0.0 \\
\hline\hline
\multicolumn{4}{c} {SPS5}\\ \hline
\hline
SI (proton)    &   6.97E-11     & 4.33E-11   & 4.6 \\
SI (neutron)  &   6.95E-11     &    4.31E-11 & 4.0 \\
SD(proton)    &   4.05E-8    &  4.09E-8  &  1.0 \\
SD (neutron) &    4.77E-7   &  4.77E-7 & 0.0  \\
\hline\hline
\multicolumn{4}{c} {SPS9}\\ \hline \hline
SI (proton)    &  5.01E-11      &   5.03E-11  & 0.3\\
SI (neutron)  &  5.00E-11     &  4.98E-11 & 0.3 \\
SD(proton)    &   1.11E-6    &  1.12E-6  & 0.6 \\
SD (neutron) &   1.72E-6    &   1.72E-6 & 0.0\\
\hline
\hline
\end{tabular}
  \caption{{\maddm{}} and {{MicrOMEGAs}} comparison for DM-nucleon cross sections in MSSM.}
  \label{MSSMtable}
\end{table}
\subsection{Recoil Rates}

\begin{figure}[t]
\centerline{
\hspace{0.6cm}
\includegraphics[scale=0.43]{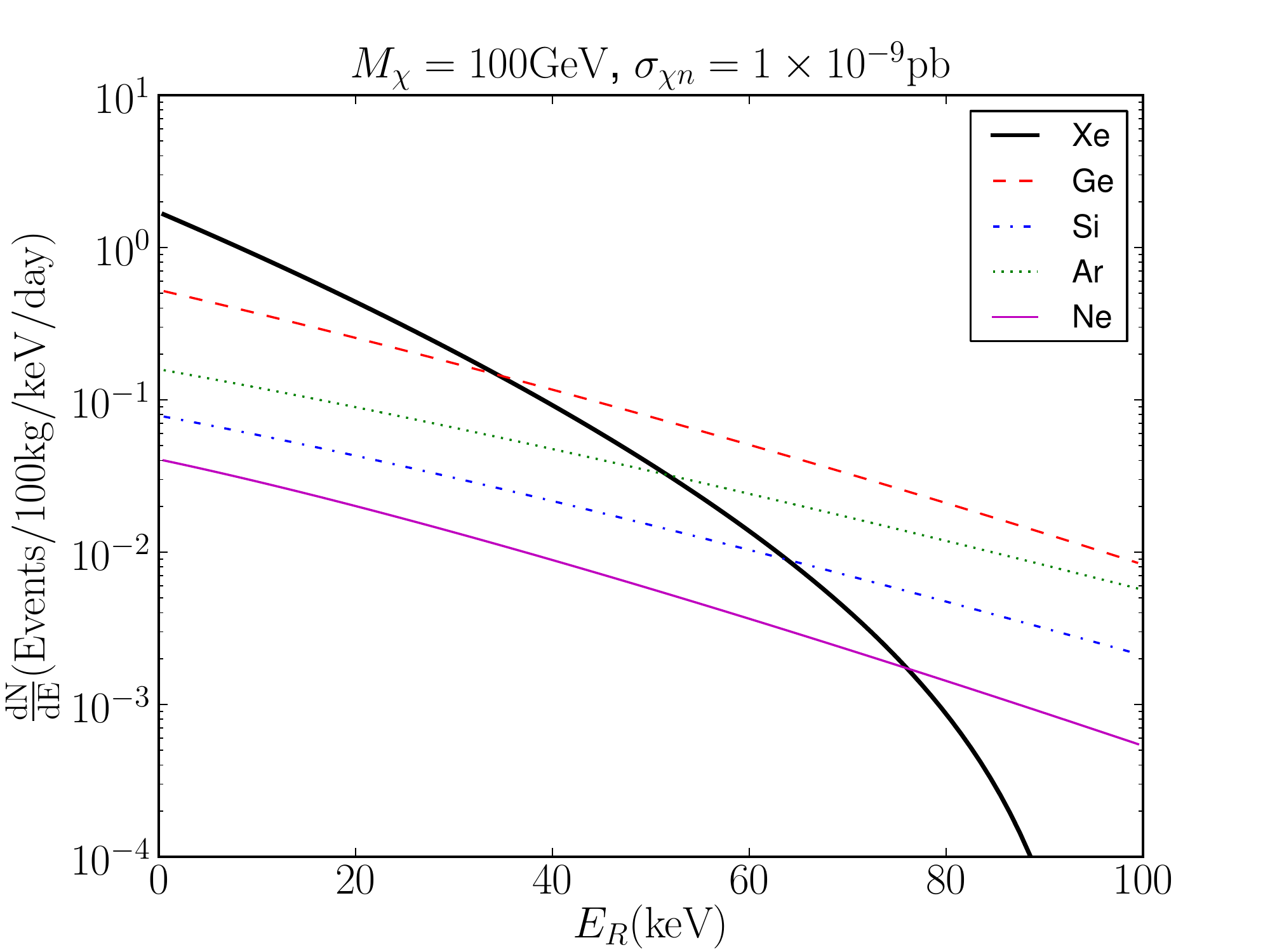}
\hspace{-0.6cm}
\includegraphics[scale=0.43]{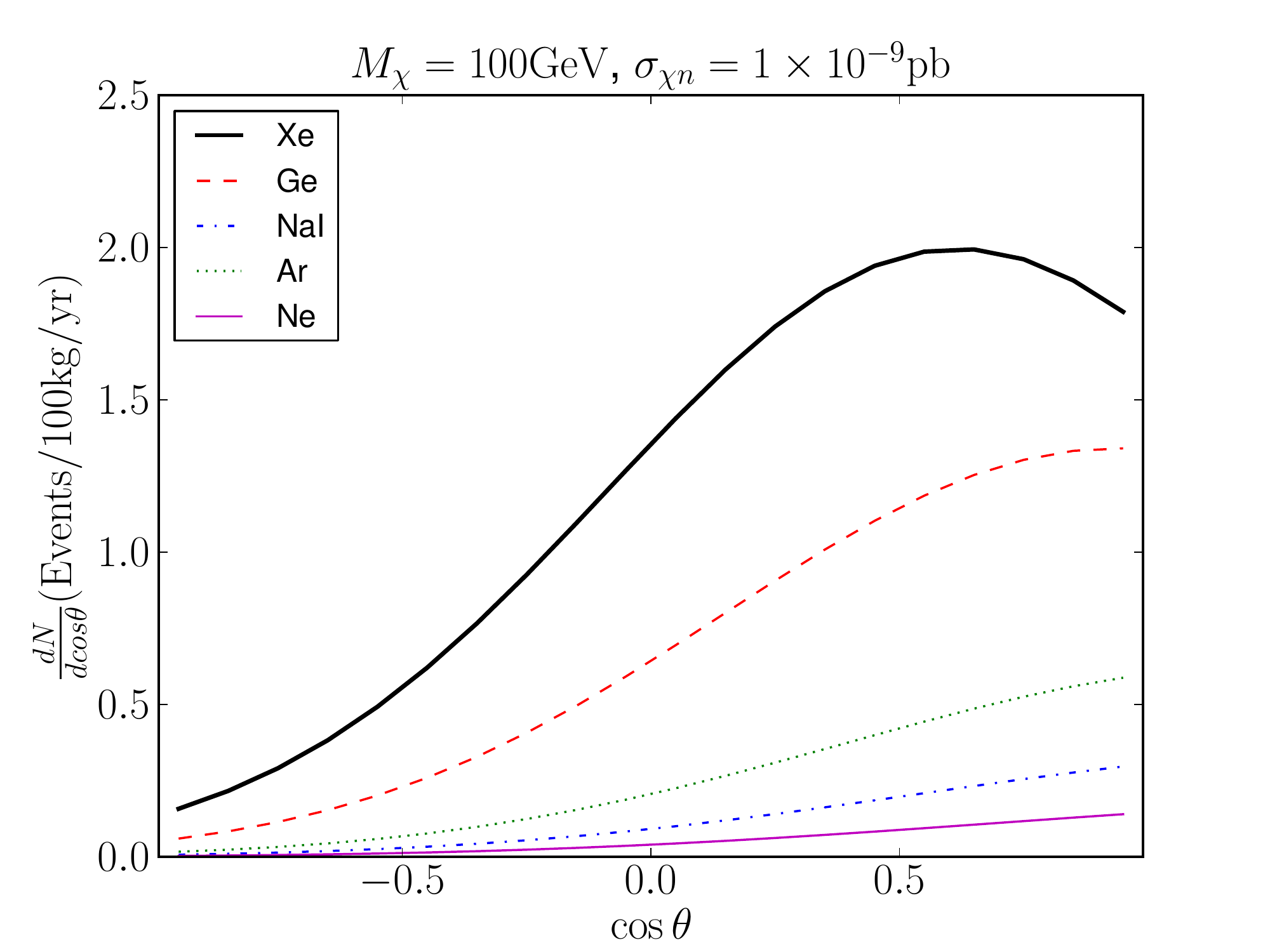} }
\caption{Nuclear recoil energy (left panel) and angular (right panel) distributions for spin-independent interactions for different materials, assuming a 100 kg detector measuring events over one year for a DM mass of 100 GeV and DM-nucleon cross-section of $1\times 10^{-9}$ pb.  \label{dndetarek}}
\end{figure}

Upon validating the DM-nucleon scattering cross section results from \maddm{},  we proceed to the recoil rates for DM scattering off a target nucleus.  
We begin with a simple, model independent validation of the recoil rate calculation, where we simply assume that the DM-nucleon cross section $\sigma_{\chi n} = 10^9$ pb, chosen for the purpose of comparison with the results from Ref. \cite{Saab:2014lda}.
To reproduce the SI recoil rates as a function of energy/angle as in Ref. \cite{Saab:2014lda}, we employ the differential recoil spectrum of Eq.~\eqref{doubleratefin}, integrated over time and angle/energy. Fig. \ref{dndetarek} shows the spin-independent recoil rates as a function of recoil energy (left) and recoil angle (right). We find that both distributions are in a very good agreement with the results found in Ref. \cite{Saab:2014lda}, over a wide range of target materials.

\begin{figure}[t]
\centerline{
 \includegraphics[scale=0.48]{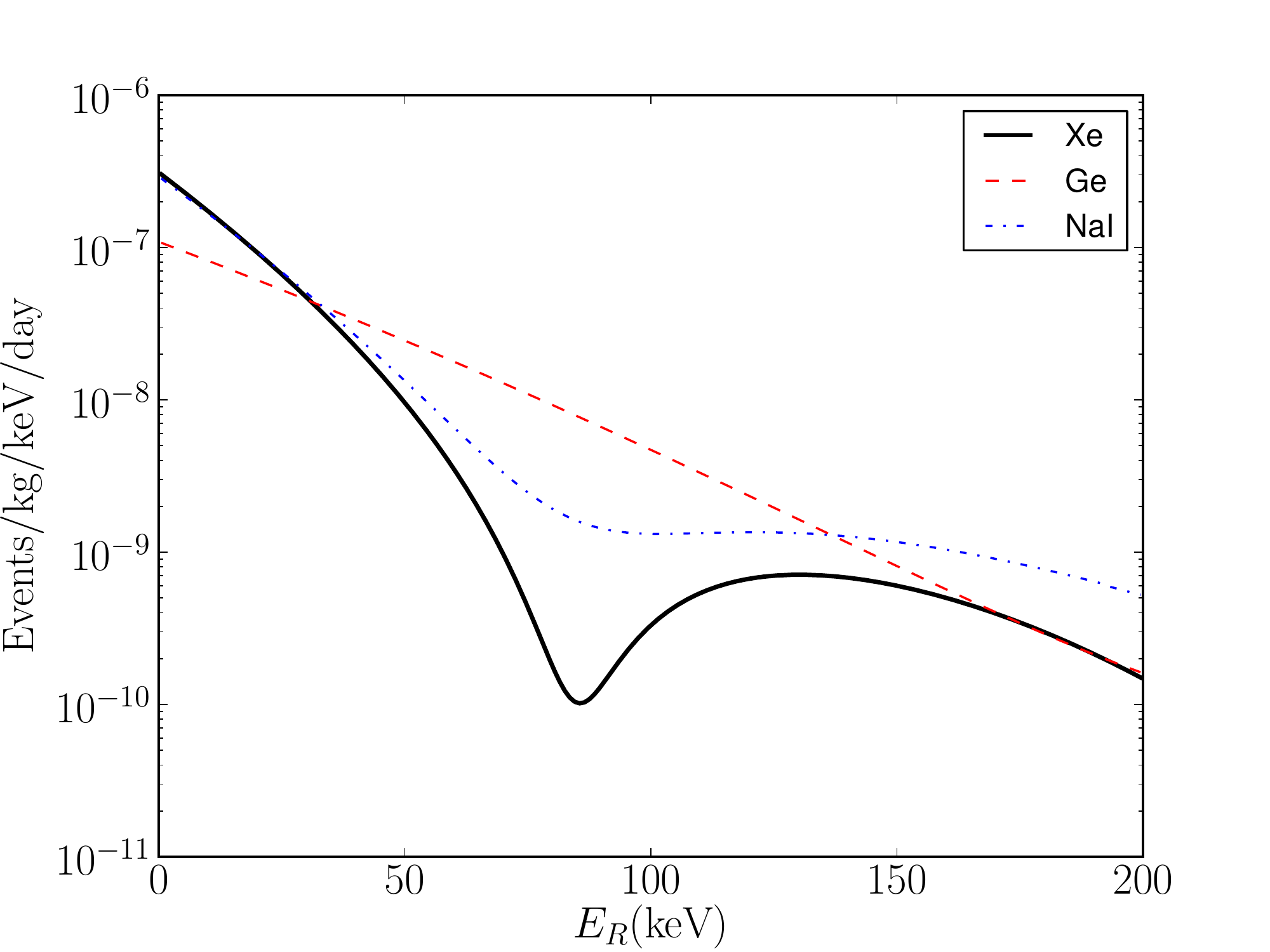} }
  \caption{Nuclear recoil energy distributions as a function of recoil energy for Xenon, Germanium and NaI. These are the rates for the UED model assuming a mass of 1000 GeV for the DM ($\gamma_{1}$) with the DM-nucleon scattering cross-sections for both spin-dependent and spin-independent for $\Delta = 15\%$ as illustrated in Fig.~\ref{ued}.}
 \label{dndets}
\end{figure}

As a next validation, we check the recoil rates in UED model following the procedure described in Ref. \cite{Servant:2002hb}, which 
shows sum of SI and SD recoil rates. The spin-dependent recoil rates are sensitive to numerical values of various quantities such as magnetic moments and parametrization of form factors. 
We use those values quoted in the references that are cited in Ref. \cite{Servant:2002hb}.
Fig. \ref{dndets} shows nuclear recoil energy distributions as a function of recoil energy for Xenon, Germanium and NaI. 
KK photon mass is chosen to be 1000 GeV with the DM-nucleon scattering cross-sections for both spin-dependent and spin-independent for $\Delta = 15\%$ as illustrated in Fig.~\ref{ued}.
Despite the minor differences in the spin-dependent recoil rates, we find that Fig. \ref{dndets} shows a good agreement between \maddm{} and the calculations in Ref. \cite{Servant:2002hb}. 

\begin{figure}[t]
\centerline{}
 \includegraphics[scale=0.48]{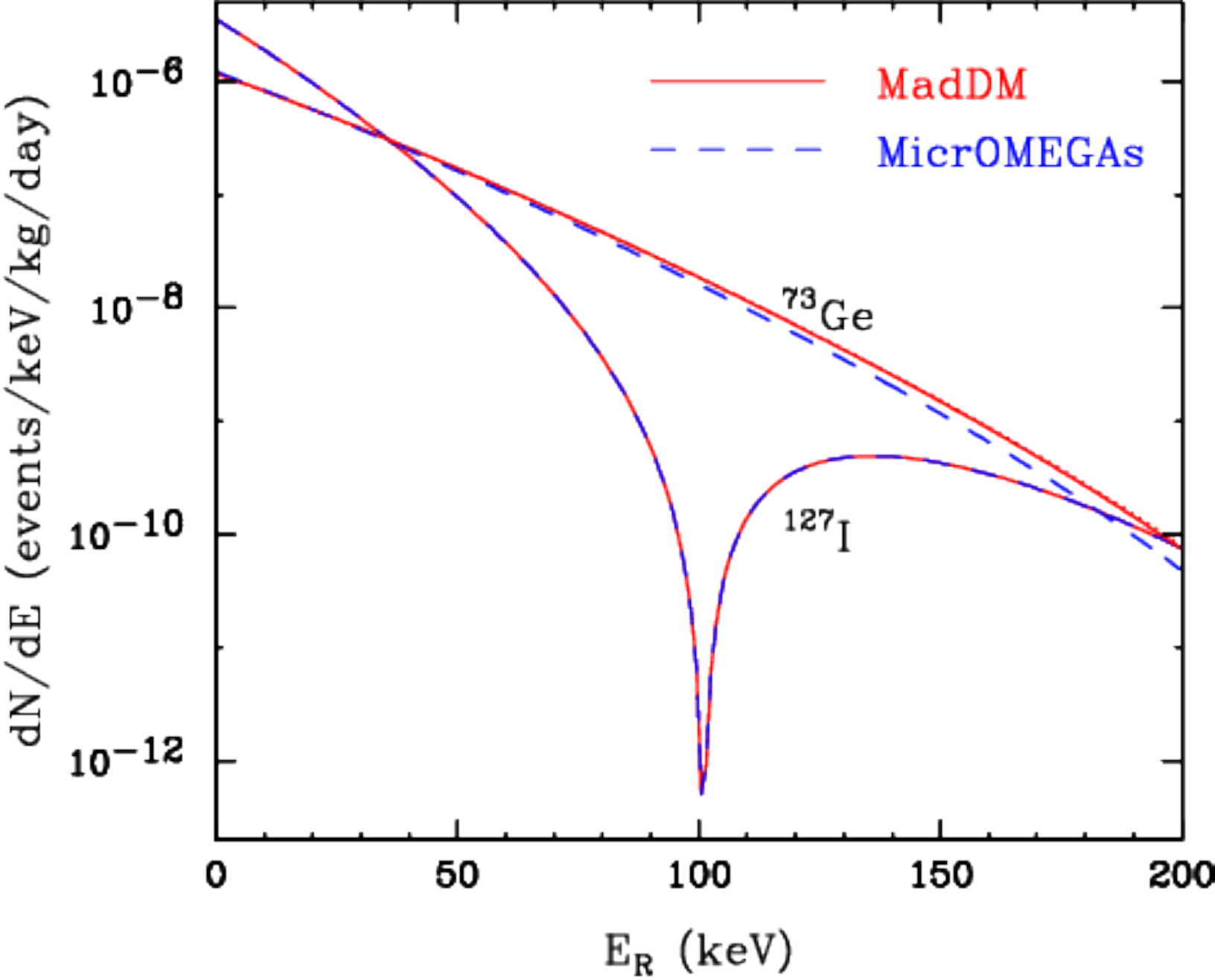}
\hspace{0.25cm}
 \includegraphics[scale=0.48]{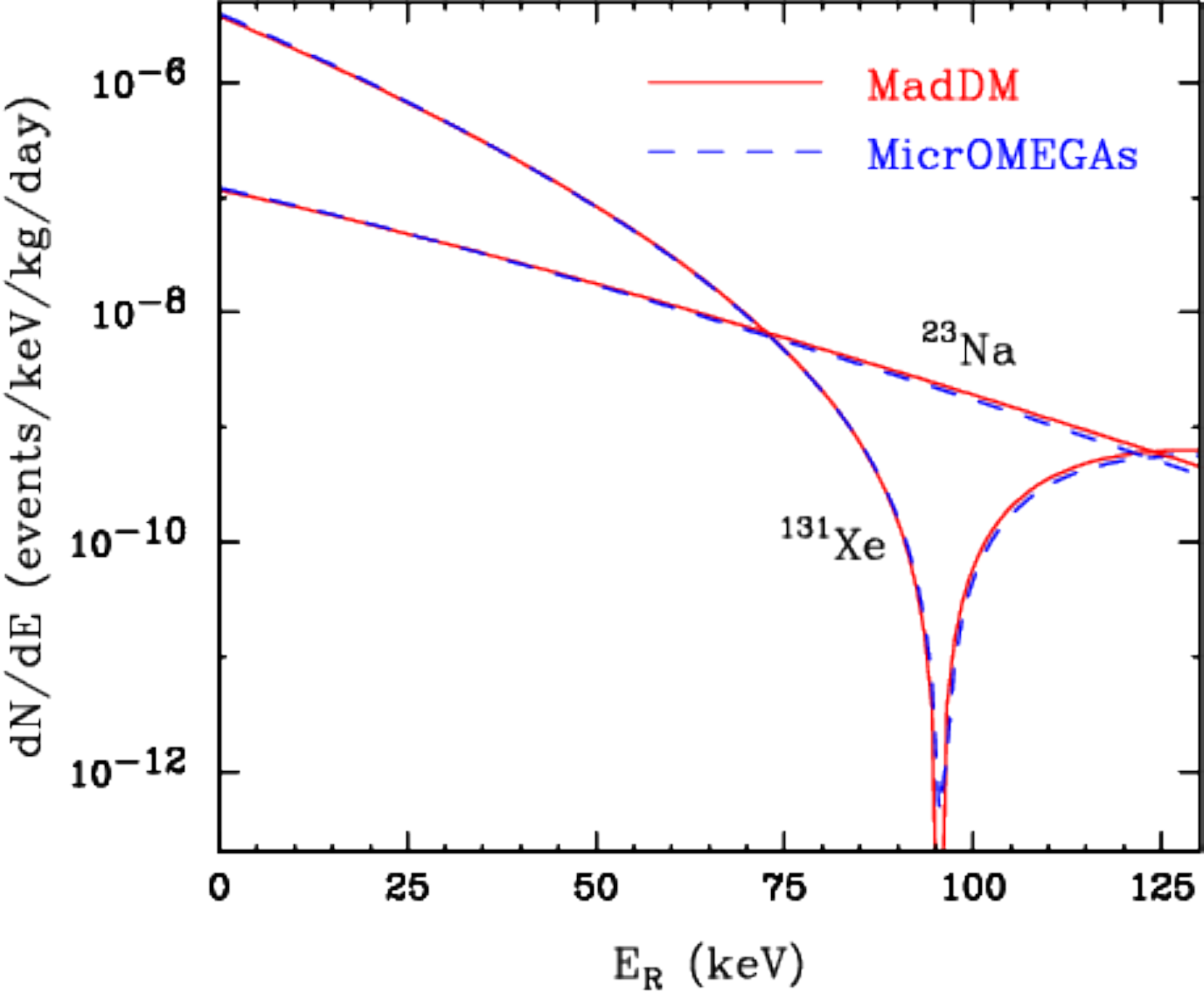}
\caption{Comparison of recoil energy distribution for I and Ge (left panel) and for Xe and Na (right panel) for a Higgs portal scalar dark matter model.}
 \label{energyrates}
\end{figure}

Furthermore, we compare the differential energy rates obtained from MicrOMEGAs with those obtained from \maddm{} for a Higgs portal scalar dark matter model for 4 different materials: Xenon, Iodine, Germanium and Sodium. We consider only the most abundant isotopes of these materials as is assumed in MicrOMEGAs. 
SD cross section vanishes in this case since DM is a scalar. 
Fig. \ref{energyrates} illustrates the comparison between MicrOMEGAs and \maddm{} for the differential energy rates as a function of recoil energy in keVnr, while Table \ref{eventscomp} shows several numerical comparisons of the total expected number of events for a 1 kg-day normalization. We find an excellent agreement between the two codes over a wide range of recoil energies and target materials. 

\begin{table}[t!]
\centering
\begin{tabular}{|c|c|c|} 
\hline
    & ~MicrOMEGAs~ & ~~~~ \maddm{} ~~~~ \\ \hline
\hline
$^{131}$Xe & $5.41\times 10^{-5}$& $5.42\times 10^{-5}$ \\ \hline
$^{73}$Ge & $ 2.99\times 10^{-5}$ & $3.07\times 10^{-5}$ \\ \hline
$^{127}$I & $5.28\times 10^{-5}$ & $5.30 \times 10^{-5}$ \\ \hline
$^{23}$Na & $3.09\times 10^{-6}$ & $3.12\times 10^{-6}$\\ \hline
\end{tabular}
\caption{\label{eventscomp} Total number of events for a Higgs portal dark matter model with 100 GeV mass. We compare the number of events as obtained from MicrOMEGAS and \maddm{} for a 1kg detector measuring events over a day. }
\end{table} 

\subsection{LUX exclusion bound}

As a final validation of the \maddm{} code, we attempt to reproduce the exclusion limit on the DM-nucleon cross-section as a function of DM mass similar to the LUX 2013 experimental results \cite{Faham:2014hza}. For this purpose, we assume the efficiency function of nuclear recoils displayed in the black curve of Fig. 1 in Ref. \cite{Akerib:2013tjd}.  

\begin{figure}[t]
\centerline{
\includegraphics[scale=0.5]{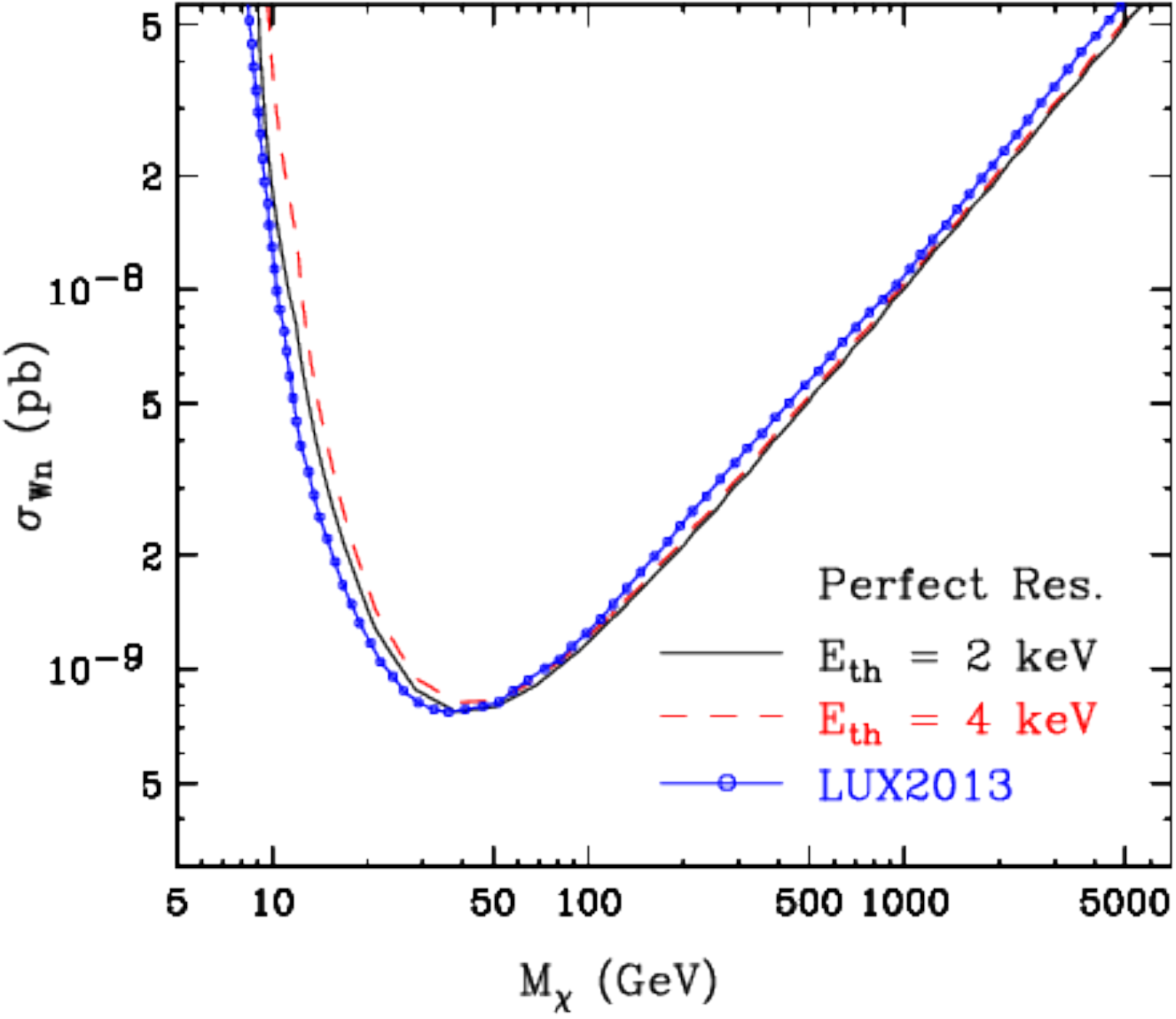}
\hspace{0.4cm}
\includegraphics[scale=0.5]{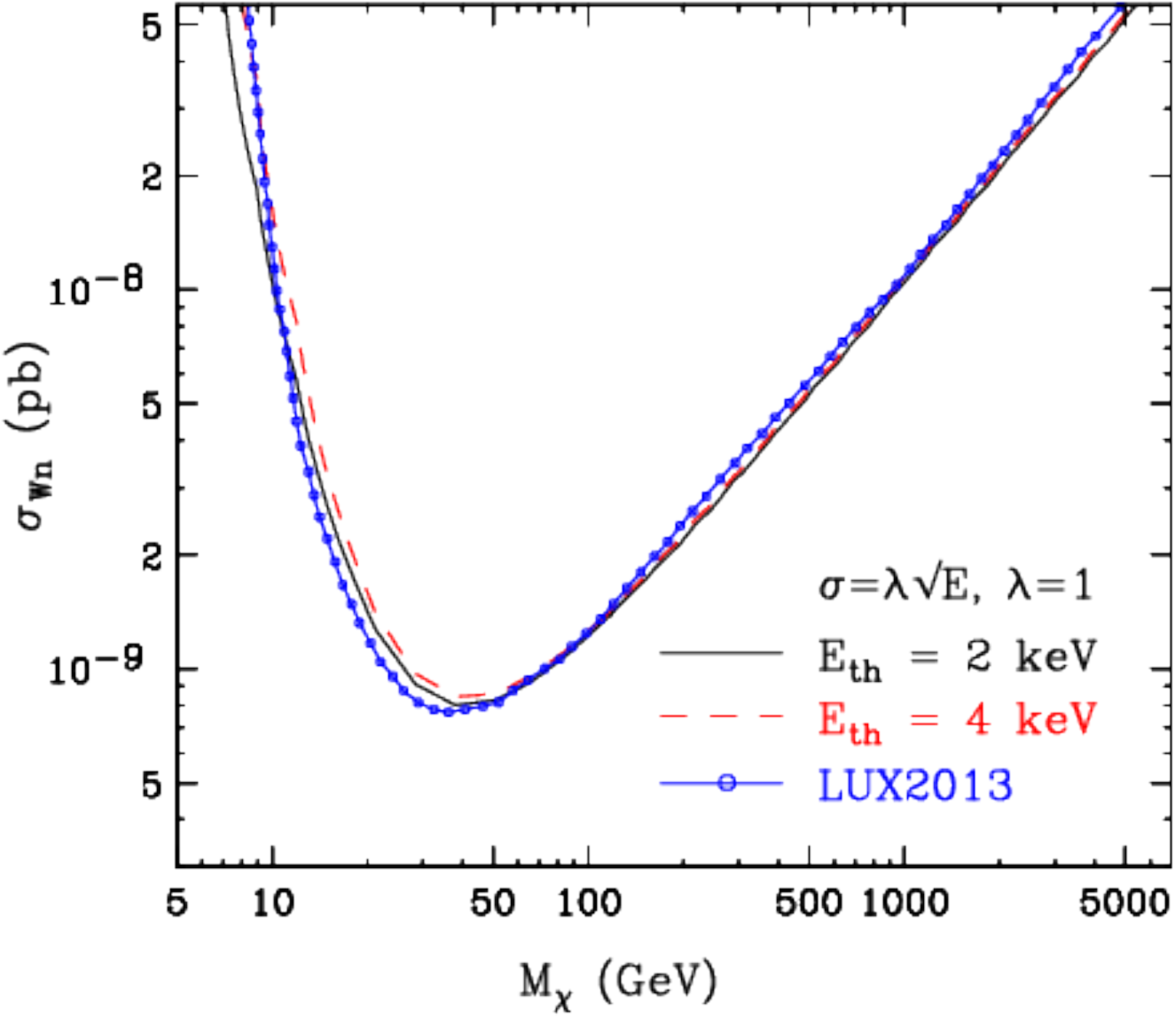} }
\caption{ 90\% confidence limits on the spin-independent DM-nucleon scattering cross-section (in picobarns) 
for an unsmeared energy distribution  (left panel) and the smeared distribution with $\lambda = 1$ (right panel). 
Limits are obtained from \maddm{} for 2 keV (black solid) and 4 keV (red dashed) and 
LUX limits are shown in blue curve with circular data points.}
\label{lux_rep}
\end{figure}

Fig. \ref{lux_rep} shows the results from \maddm{} for different energy threshold cuts as compared to the data reported by the LUX experiment. We present the contours assuming 2.3 events, coinciding with the number of events at 90\% confidence as required by the Feldman-Cousins confidence intervals. We find a good match between the LUX data and limits from \maddm{}. As we could not obtain information on what value of the energy threshold cut is used in the LUX limit, we considered different values of threshold cuts. As illustrated in Fig. \ref{lux_rep}, the threshold cuts only impact the lower mass side since a higher threshold cut reduces the statistics for a lower DM mass. 

Note that, as described in the previous sections, the exclusion curves in Fig. \ref{lux_rep} can be obtained in \maddm{} by using \verb|LUX_Exclusion| routine found in the test routines part in \verb|maddm.f|. The routine multiplies $\frac{dR}{dE}$ by the efficiency obtained from Ref. \cite{Akerib:2013tjd}, which is is then weighted by a 50 \% acceptance rate for nuclear recoils as stated in the LUX analysis.  From the recoil spectrum weighted by the efficiency and the acceptance rate, the function then calculates the total number of expected events. 
The default value for the detector efficiency is 100\%, and can be easily replaced by a user defined function.

%


\section{Conclusions}\label{sec:conclusions}
The identity of dark matter is one of the most profound mysteries in particle physics, astrophysics, and cosmology.
Recent data from gamma rays, supernovae luminosities, cosmic microwave anisotropies, and galactic rotation curves all point consistently 
to the existence of dark matter with $\sim 5$ times more abundance compared to ordinary matter.
At the same time, all known particles are excluded as possible dark matter candidates, 
making the dark matter problem perhaps the most pressing motivation for physics beyond the Standard Model.
Little is currently known about the mass scale of dark matter, suggesting that discovery and characterization of DM will likely require a synergistic approach including well-balanced programs in direct detection, indirect detection,  particle colliders and astrophysical probes. 
 
In order to efficiently combine results from various dark matter searches sparked a demand for a new generation of numerical tools. \maddm{} is an on-going effort to bridge DM collider phenomenology with astro-physics and cosmology of DM, with the ultimate goal to provide an ``all in one'' dark matter phenomenology package 
which can be easily incorporated into the future dark matter searches at the LHC.

In our current work, we presented \maddm{} 2.0, which includes direct detection of dark matter in a generic UFO model.
The code computes the total DM-nucleus scattering rate, recoil energy spectrum and angular distributions of dark matter elastic scattering off nuclei for various target materials, 
including user-provided detector resolution and detector efficiency. \maddm{} v.2.0 follows the ``easy to use, easy to install'' philosophy of the previous version, featuring great flexibility so that users can replace the existing modules with their own implementation or link \maddm{} functionality with other existing tools. 
We have performed detailed valuation of MadDM against MicrOMEGAs and private codes for various benchmark dark matter models and found excellent agreements with existing tools and literature. 
 
In addition to relic abundance and direct detection, the future renditions of the \maddm{} code will also provide ability to calculate cosmic-ray fluxes from DM annihilation, including gamma rays, cosmic electrons/positrons and protons/anti-protons. Furthermore, recent improvements in aMC@NLO framework allow for computations of amplitudes for loop induced diagrams \cite{Alwall:2014hca}, giving us an opportunity to create the first automated DM tool which can calculate relic density, direct detection and indirect-detection signals originating in loop-induced processes.



\bigskip
\emph{Acknowledgments:} 
We are grateful to Valentin Hirschi for helpful discussions and advice on the \madgraph , and 
Adam Para and Jonghee Yoo for discussion on directional dark matter detection, and 
Fabio Maltoni for comments on our manuscript and encouragement on development of MadDM. 
We also thank Benjamin Fuks with questions regarding MSSM implementation in FeynRules.
We also thank organizers and participants of MC4BSM workshops for their questions and comments during meetings. 
This work is supported in part by the Belgian Federal Science Policy Office through the Interuniversity Attraction Pole P7/37, 
by the National Research Foundation of South Africa under Grant No. 88614, and by the U.S. DOE under Grant No. DE-FG02-12ER41809, 
and by Durham International Junior Research Fellowships.

\bigskip

\appendix
\section{Structure of \maddm{} v.2.0}\label{sec:app1}
\maddm{} v.2.0 follows the directory and file structure of MadDM v.1.0. The \verb|Python| module is arranged into the following simple structure:
\begin{itemize}
\item \verb|EffOperators:| A folder containing UFO files for effective operators, which are used to extract the low energy coefficients of  the $\chi q \rightarrow  \chi q$ matrix elements. The UFO files were generated using the FeynRules 2.0 convention (see Section \ref{UFOconvention} for more details.). 
\item \verb|ExpData:| all experimental data used by \maddm{} is stored in this folder. Currently, only the LUX exclusion bound for the spin-independent DM-nucleon cross section is included. Future versions of \maddm{} will include a more elaborate database of experimental bounds.
\item All user defined projects are automatically output into the  \verb|Projects| folder. All numerical calculations and result output is written into the \verb|Projects| folder. 
\item \verb|Templates| contains necessary \verb|FORTRAN| routines and template files for relic density and direct detection. The structure of the \verb|Template| folder serves as a skeleton for a user defined Project.
\item \verb|MGoutput.py| contains a simplified version of the MadGraph output routines, which write the matrix elements into \verb|FORTRAN| files. The routines within \verb|MGoutput.py| will write the \verb|matrix.f| files, into the \verb|Projects| folder, but will omit writing the additional files which are not necessary for \maddm{} to run, but are ordinarily written by MadGraph output routines.
\item \verb|darkmatter.py| contains the essential part of \maddm{} \verb|Python| module. The routines within this library contain class definitions for the \verb|darkmatter| objects, as well as numerous functions which find dark matter candidates and generate the \maddm{} \verb|FORTRAN| module.
\item A self contained example of a \maddm{} calculation without the use of the user interface can be found in \verb|example.py|. The example calculation goes through the process of importing a user defined UFO model, finding the DM candidate, generating the \verb|FORTRAN| module and calculating the DM-nucleon cross section and relic density. 
\item \verb|init.py| contains the definitions and functions used by the \maddm{} user interface.  
\item \verb|maddm.py| is the main \verb|Python| executable. It will initiate the user interface defined in \verb|init.py| and lead the user through the steps of DM relic density and calculations relevant for DM-nucleon scattering.
\item \verb|param_scan_default.py| is a pre-defined parameter scanning \verb|Python| script. The file contains a skeleton of  the \verb|Python| code which will scan over DM model parameters and calculate relevant physical quantities. In order to run the parameter scanning script, a minimum of user intervention is required. All places where user input is necessary are marked, and limited to the choice of DM model parameters, number of parameter to scan over and definitions of parameter ranges. 
\end{itemize}

The \verb|FORTRAN| module in each \verb|Projects| folder contains the following 7 subdirectories.
\begin{itemize}
\item \verb|Cards:| A directory which contains \verb|param_card.dat|. The structure and placement of the parameter card follows the standard formats of the \madgraph{} framework. 
\item \verb|Source| directory contains \verb|FORTRAN| code for the DM model, as well as necessary HELAS libraries. 
\item \verb|include| directory contains various include files for the \verb|FORTRAN| code. The \verb|maddm_card.inc| which contains user defined parameters of the DM relic density and direct detection calculations is located in the \verb|include| folder. Note that any changes in the \verb|maddm_card.inc| file will only be reflected upon the re-compilation of the \maddm{} \verb|FORTRAN| module. 
\item Amplitudes for relic abundance and direct detection are saved under \verb|matrix_elements|.
\item All results (relic density, cross sections and distributions) are found in \verb|output| directory.
\item \verb|src| directory contains the \maddm{} \verb|FORTRAN| code.  The main program is located in the \verb|maddm.f| file.
\end{itemize}

All the other files in the main directory are executable and perform the following tasks.
\begin{itemize}
\item \verb|maddm.x| is the main \verb|FORTRAN| executable. Note that all output of the \verb|maddm.x| will be written into the \verb|output/maddm.out| file by default. 
\item \verb|make_plots| is a \verb|Python| script containing calls to plotting \maddm{} plotting routines. All calls are commented out by default. Note that it is necessary to install both \verb|Numpy| and \verb|Matplotlib| in order for plotting functionality to work in \maddm{}. 
\item \verb|plotting.py| is a library of user friendly plotting functions. See \verb|make_plots| for examples of how to use plotting functionality of \maddm{}. Function definitions in \verb|plotting.py| use libraries from both \verb|Numpy| and \verb|Matplotlib|. 
\end{itemize}


\section{Description of \maddm{} v.2.0 routines}
\subsection{Python Module}

We upgraded the \verb|Python| module of \maddm{}  to allow users to select which calculation they would like to perform. For this purpose we introduced three switches, which are members of \verb|darkmatter| class:

\begin{itemize}
	\item \verb|<darkmatter> _do_relic_density|: \verb|True| if the user wants to calculate dark matter relic density, \verb|False| otherwise. 
	\item \verb|<darkmatter> _do_direct_detection|:  \verb|True| if the user wants to calculate spin-dependent and spin-independent cross sections for dark matter scattering off protons and neutrons, \verb|False| otherwise. 
	\item \verb|<darkmatter> _do_directional_detection|: \verb|True| if the user wants to calculate recoil rates (as a function of energy and angle) assuming detectors made of various materials. Please note that this flag can not be set to \verb|True| if \verb|_do_direct_detection = False|.
\end{itemize}

In order to reflect the changes in the code, we have changed the name of the function which executes the numerical module of \maddm{} from \verb|CalculateRelicDensity()| to \verb|Calculate()|. The function now takes the output file name, \verb|output_file| as a parameter, specifying the path of the file where all the calculation results will be written. The default value for the output file is \verb|output/maddm.out|. Changing the output file path is particularly useful if the user wishes to parallelize a parameter scan, as it avoids the problem of many jobs attempting to read/write the same file.

The only two new relevant \verb|Python| functions in the \verb|darkmatter| class are: 

\begin{itemize}
\item \verb|darkmatter: [excluded, excluded_omegah2, excluded_dd]:| \\ \verb|is_excluded( omega_min, omega_max , si_limit_x, si_limit_y ): | This function compares the results of a DM calculation obtained with the \verb|Calculate()| function with the existing bounds on relic density and spin-independent DM-nucleon scattering cross sections. The parameters \verb|omega_min| and \verb|omega_max| represent the minimum and maximum $\Omega h^2$ for which the parameter point is not ruled out while the \verb|si_limit_x| and \verb|si_limit_y| are one dimensional arrays of numbers representing the $x$ and $y$ values of the spin-independent cross section exclusion limit. The default limit implemented in \maddm{} is the LUX bound, which can be found in a text file inside the \verb|ExpData| subdirectory of \maddm{}. \textbf{The function assumes that the DM-nucleon cross section values are given in pb.} The function returns an array of three boolean variables. \verb|excluded| signals whether the model point is excluded overall, while \verb|excluded_omegah2| and \verb|excluded_dd| signal whether the model point is excluded by the individual constraint. This function should be called only after \verb|Calculate()| has been executed.

\item \verb|darkmatter: None: GenerateDiagramsDirDetect():| The routine generates all $\chi q \rightarrow \chi q$ diagrams, where $q = u,d,s,c,b,t$ quarks, as well as $\chi \bar{q} \rightarrow \chi \bar{q}$ diagrams and extracts the interference terms relevant for computing the low energy effective operators for $\chi$-nucleon scattering. The function takes no values and returns no values. 
\end{itemize}

Simply calling \verb|GenerateDiagramsDirDetect()| will automatically set \verb|_do_direct_detection| to \verb|True| and result in a calculation which will go as far as $\chi$-nucleon spin-independent and spin-dependent cross sections. The user can proceed to calculate the differential recoil rates and simulate dark matter events scattering off nuclei by also setting the flag \verb|_do_directional_detection = True|. 

\subsection{FORTRAN Module}

In the \verb|FORTRAN| module, we added a library of function definitions in \verb|direct_detection.f| and \\ \verb|directional_detection.f| files inside the \verb|src| directory, which calculate the direct detection nucleon cross sections, differential recoil rates and other relevant quantities. The relevant function definitions are:
\begin{itemize}
\item \verb|sigma_nucleon(nucleon):| The function calculates the numerical value of a cross section for  dark matter particle scattering off a nucleon at 0 momentum transfer. \verb|nucleon=1,0| refers to a proton/neutron respectively. The function reads in the nucleon form factors as defined in \verb|maddm_card.inc| (see below for more detail). 

  \item \verb|FHelm(Er, A):| This function calculates the Helm form factor as given in Eq. \eqref{formfac}. The function reads the recoil energy in keVnr (\verb|Er|), and the mass number of the target atom (\verb|A|) as input parameters.
  
 \item \verb|FWS(Er, A):| This function calculates the Wood-Saxon form factor. The function reads the recoil energy in keVnr (\verb|Er|), and the mass number of the target atom (\verb|A|) as input parameters.
    
 \item \verb|FWSmass(m_DM, A):| This function calculates the Wood-Saxon form factor. The function reads the WIMP mass in GeV (\verb|m_DM|), and the mass number of the target atom (\verb|A|) as input parameters.
  
  \item \verb|coefficients(mater,coeff_s00,coeff_s01,coeff_s11):| This routine outputs the coefficients for the polynomial fits to the nuclear structure functions $S_{ij}$ for the spin-dependent form factor. These are referenced in table \ref{targetsd}. The input is the material \verb|mater| which is chosen in \verb|maddm_card.inc|.
  
  \item \verb|Spin_matrix(mater, J, avg_sp, avg_sn):| Provides the spin \verb|J| and nuclear magnetic moments \verb|avg_sp| and \verb|avg_sn| for the different target nuclei. \verb|mater| is the material chosen in \verb|maddm_card.inc|. 

  \item \verb|Structure_Func(Er, A, mater, s00, s01, s11):| Uses the \verb|coefficients| routine to calculate the structure functions (\verb|s00|, \verb|s01| and \verb|s11|) that are used in the spin-dependent nuclear form factor. The input is the recoil energy \verb|Er| in keVnr, the atomic mass \verb|A| and the target material \verb|mater|.
  
  \item \verb|FormSD(Er, A, mater):| Calculates the spin-dependent form factor as a function of recoil energy in keVnr  from the structure functions in \verb|Structure_Func| according to Eq.~\eqref{formsd}.
  
  \item \verb|FormSDmass(m_DM, A, mater):| Calculates the spin-dependent form factor as a function of WIMP mass in GeV from the structure functions in \verb|Structure_Func| according to Eq.~\eqref{formsd}.
  
  \item \verb|VE(days):| This function calculates the velocity of the earth (in km/s) relative to the galactic center as a function of time in days according to appendix B in \cite{Lewin:1995rx}. The function takes into account the proper motion of the solar system with respect to the surrounding stars, the motion of the Earth around the Sun and provides galactic coordinates of the Earth as it moves each day on its orbit.   The input parameter \verb|days| represents the day of the year relative to December $31^{\mathrm{st}}$ of the calendar year. 
  
  \item \verb|d2RdEdcostheta|\\ \verb|(Er, costheta, sigmawn0SI, sigmawp0SI, sigmawn0SD, sigmawp0SD, M_dm, ve):|\\ This routine calculates the recoil rates as a function of energy and angle 
  on a particular day of measurement, based on Eq.  \eqref{doubleratefin}. 
The \verb|maddm_card.inc| file contains numerical values of necessary quantities such as 
 the target material, size of the detector (in $\rm kg$), the velocity of the earth (calculated inside $\rm VE$) , the most probable DM velocity $v_{0}$, the escape velocity of the DM from the galactic Halo $\rm v_{esc}$, the minimum DM velocity required for recoil $\rm v_{min}$  all in units of km/s,  and the local halo density $\rm \rho_{0}$, in $\rm GeV c^{-2} cm^{-3}$. 
  The variables \verb|sigmawp0SI| and \verb|sigmawn0SI| represent the spin-independent cross sections, while \verb|sigmawp0SD| and \verb|sigmawn0SD| are the spin-dependent cross-sections for DM scattering off protons and neutrons respectively in units of $\rm pb$. \verb|M_dm| and \verb|ve| are the dark matter mass in $\rm GeV$ and the Earth's velocity relative to the Galactic center respectively. 
   
  \item \verb|target_material(material, A, Ab, Z, n):| This subroutine provides the double differential module \verb|d2RdEdcostheta| with the correct target material the user has chosen inside \verb|maddm_card.inc|. The input is an integer \verb|material| provided through a common block from the \verb|maddm_card.inc| file, while the output is an array of mass numbers (\verb|A|) for the target atoms, their respective abundances (\verb|Ab|), the atomic number  (\verb|Z|) as well as the length of the mass number arrays (\verb|n|). These are listed in Table \ref{targetsi}. For instance, consider that the target material has been set to 1 (Xenon, which is default) in the  \verb|maddm_card.inc| file. Then \verb|target_material| will output an array of 9 elements for the mass number of the 9 most stable Xenon isotopes and their corresponding abundance information. It will also provide $\rm Z = 54$.  
    
  \item \verb|GaussmearAngle(uaevents, smaevents):| This function smears a histogram of theoretical distributions (\verb|uaevents|) using a Gaussian distribution.
The angular resolution is set inside \verb|maddm_card.inc| by changing the \verb|sig_theta| parameter \cite{Mohlabeng:2015efa}. The default angular resolution is $30^\circ$. 
  
  \item \verb|GaussmearEnergy(uevents, smevents):|  This module applies energy smearing to the distribution of angle smeared events (\verb|uevents|) using a Gaussian resolution function with uncertainty $\sigma_E = \lambda \sqrt{E_R}$, where $E_R$ is the recoil energy in keV and $\lambda$ is a constant which determines the level of smearing. The default value is $ \lambda = 1$ inside \verb|maddm_card.inc|. The smearing function can convert low energy recoil events into negative energies. An energy threshold cut alleviates this problem. 
 
  \item \verb|Theory_Simulation|\\ \verb|(mass, sigma_wnSI, sigma_wpSI, sigma_wnSD, sigma_wpSD, Theory):| This subroutine uses the dark matter mass in GeV (\verb|mass|) and DM-Nucleon cross-sections for the spin-independent (\verb| sigma_wpSI| for proton, \verb|sigma_wnSI| for neutron ) and spin-dependent (\verb| sigma_wpSD| for proton, \verb|sigma_wnSD| for neutron ) contributions to simulate the theoretical double differential distribution (with respect to recoil energy and recoil angle) and outputs the result as an array named \verb|Theory| representing the distribution of unsmeared recoil events. In this routine, the double differential rate is approximated at the center of each energy-angle-day bin. 
  
  \item \verb|eff(x):| Function that interpolates the data points from LUX for the nuclear recoil efficiency. The input \verb|x| is the recoil energy (in keV) and the output is the detector efficiency at that energy.
  
  \item \verb|efficiency(x, flag):| Provides the DM detection efficiency as a function of recoil energy. The variable \verb|x| represents the recoil energy in keV and flag is an integer variable which, when chosen as \verb|0|, gives a constant $100\%$ detection efficiency. If \verb|flag| is \verb|1|, the efficiency is $50\%$ and if \verb|flag| is \verb|2|, the efficiency is that obtained from the function \verb|eff|, which is the nuclear recoil detection efficiency from the LUX experiment in Fig.~1 of Ref. \cite{Akerib:2013tjd}. For a conservative estimate we extend and interpolate the efficiency to 100 keV. New detector efficiency may be implemented by user in place of this function definition.
    
  \item \verb|directional_detection|\\ \verb|(mass, sigma_wnSI, sigma_wpSI, sigma_wnSD, sigma_wpSD, flag, N_events):| This routine calculates the differential distributions of dark matter recoil rates given the spin-independent (\verb| sigma_wpSI| for proton, \verb|sigma_wnSI| for neutron) and spin-dependent \\(\verb| sigma_wpSD| for proton, \verb|sigma_wnSD| for neutron) cross sections in pb as well as the DM mass in GeV (\verb|mass|).  The subroutine calls the \verb|theory_simulation| routine which performs the event distribution simulation. If the user specified \verb|smearing = .true.| in the \verb|maddm_card.inc| file the function will incorporate detector smearing into the resulting distributions. The function calculates the double differential distribution (as a function of energy and angle), and integrates over angle and time to obtain the differential energy distribution ($\frac{dN}{dE}$). The energy distribution is weighted with the detector efficiency function \verb|efficiency|. Furthermore, the function also integrates over energy and day to obtain the differential angular distribution ($\frac{dN}{dcos \theta}$ ) as well as the total rate ($\frac{dN}{dt} = R$) after integrating over energy and angle.  All the distributions are written to separate files in the user defined project \verb|output| directory. The total number of events for a 1 ton detector over the course of 1 year is stored in the variable \verb|N_events|. The integer variable \verb|flag| is used to choose the type of efficiency to be used.

 \end{itemize}
In addition to functions which calculate quantities relevant for DM-nucleon scattering, we added a number of important variables to the \verb|FORTRAN| module, in the \verb|maddm_card.inc| file inside the \verb|include| directory of the project folder \footnote{Please note that one must recompile the numerical code by typing make in the project folder in order for any changes to the maddm\_card.inc file to take effect properly.} :

\begin{itemize}
\item  \verb|do_relic_density|: A logical flag which determines whether a relic density calculation should be performed. It can be either \verb|.true.| or \verb|.false.| If users have not generated the \verb|FORTRAN| module with the relic density calculation  (\textit{i.e.} there is no call to \verb|GenerateDiagrams()| in the \verb|Python| module), setting this flag to \verb|.true.| will result in an error.
\item \verb|do_direct_detection|: A logical flag which determines whether a direct detection calculation should be performed. It can be either \verb|.true.| or \verb|.false.| If users have not generated the \verb|FORTRAN| module with the direct detection calculation (\textit{i.e.} there is no call to \verb|GenerateDiagramsDirDetect()| in the \verb|Python| module), setting this flag to \verb|.true.| will result in an error.
\item \verb|do_directional_detection|: A logical flag which determines whether to calculate differential distributions for recoil rates (both energy and angle dependence). It can be either \verb|.true.| or \verb|.false.| If users have not generated the \verb|FORTRAN| module with the direct detection calculation (\textit{i.e.} there is no call to \verb|GenerateDiagramsDirDetect()| in the \verb|Python| module), setting this flag to \verb|.true.| will result in an error. In addition, one can not set to \verb|do_directional_detection = .true.| if \verb|do_direct_detection = .false.|\,. 
\item \verb|SPu... SPg, SNu... SNg:| Values for the \textbf{scalar} nucleon form factors for the proton (\verb|SPx|) and neutron (\verb|SNx|). The nucleon constituent ($u, d, ... g$) is labeled by the last letter in the variable name. 
\item \verb|VPu, VPd,  VNu, VNd:| Values for the \textbf{vector} nucleon form factors for the proton (\verb|VPx|) and neutron (\verb|VNx|). The nucleon constituent ($u, d, ... g$) is labeled by the last letter in the variable name.
\item \verb|AVPu... AVPs,  AVNu... AVNs:| Values for the \textbf{axial-vector} nucleon form factors for the proton (\verb|AVPx|) and neutron (\verb|AVNx|). The nucleon constituent ($u, d, ... g$) is labeled by the last letter in the variable name.
\item \verb|SigPu... SigPs,  SigNu... SigNs:| Values for the \textbf{tensor} nucleon form factors for the proton (\verb|SigPx|) and neutron (\verb|SigNx|). The nucleon constituent ($u, d, ... g$) is labeled by the last letter in the variable name.
\item\verb|material:| Variable which selects the detector material. It can take values 1-13 for the following implemented materials: 1 - $\rm Xe$,  2 - $\rm Ge$, 3 - $\rm Si$, 4 - $\rm Ar$, 5 -  $\rm Ne$, 6 - $\rm Na$, 7 - $\rm I$, 8 - $\rm C$, 9 - $\rm F$, 10 - $\rm S$. For compound materials set \verb|mater_comp| as \verb|.true.| in \verb|maddm_card.inc|, 11 - $\rm NaI$, 12 - $\rm CF_{4}$ and 13 - $\rm CS_{2}$.
\item\verb|vMP:| Most probable speed for dark matter particles in the galactic halo in km/s, assuming the Maxwell-Boltzmann velocity distribution. The default value of \verb|vMP| is $v_0 =220$ km/s. 
\item\verb|vescape: | Escape velocity for dark matter in the galactic halo in km/s, the default value is $v_{\rm esc}=650$ km/s.
\item\verb|rhoDM:| Mass density of dark matter in the halo in $\GeV c^{-2}\mathrm{cm}^{-3}$, default value is 0.3 $\GeV c^{-2}\mathrm{cm}^{-3}$.
\item\verb|detector_size:| Mass of the dark matter detector target in kg, the default value is 1000 kg.
\item\verb|En_threshold: | The recoil energy threshold cut (in keVnr). 
\item\verb|lambd :| Energy resolution parameter defined in $\sigma_{E} = \lambda \sqrt{E}$ with $\lambda=1$ as default.
\item\verb|sig_theta: | Angular resolution of the detector (in degrees) with default value of $\sigma_{\theta}=30^\circ$. 
\item\verb|smearing: | A logical flag which determines whether the results should be smeared by detector effects. If set to \verb|.true.| it will smear both the recoil energy and angle. 
\end{itemize}
For the purpose of dark matter scattering event simulations, we also incorporated a number of parameters which are used as inputs for the nuclear recoil scattering simulation:
\begin{itemize}
\item\verb|En_min, En_max: | Minimum and maximum recoil energy in keVnr. 
\item\verb|cos_min, cos_max: | Minimum and maximum cosine of the recoil angle. 
\item\verb|day_min, day_max: | Starting and ending day of the experiment relative to December $31^{\rm st}$.
\item\verb|Energy_bins, cos_theta_bins, day_bins: | The number of bins in histograms for scattering rate as a function of recoil energy, cosine angle and day of experiment respectively.
\end{itemize}

\subsection{Test routines}
For debugging purposes we also provide a set of test routines and functions which can be used in \verb|maddm.f| and are commented out by default. Simply un-commenting the calls and re-compiling the \verb|FORTRAN| module will make the test functions accessible. The list of test function includes the following:
\begin{itemize}
\item \verb|d2RdEdcos_test|\\ \verb|(Er, costheta, sigmawn0SI, sigmawp0SI, sigmawn0SD, sigmawp0SD, M_dm, day):| \\ This subroutine uses the  \verb|d2RdEdcostheta| routine described in the previous section to print out all the ingredients that go into the double differential function, for consistency checking. The input variables are defined in  \verb|d2RdEdcostheta| above. The output is printed on the screen and includes: DM mass, the spin-dependent and spin-independent cross-sections, the target material used as well as its mass numbers and the isotope abundances, the detector size and other variables that factor into Eq.~\eqref{doubleratefin}.
\item \verb|V_E_test(day, num_array):| This routine provides the velocity of the Earth through the galactic halo on a specific day relative to December $31^{\rm st}$. \verb|num_array| is an integer flag. When set to \verb|0|, \verb|V_E_test| prints one number representing the velocity on the day specified by \verb|day|. However when  \verb|num_array| is set to \verb|1|, the routine outputs an array of days and velocities (in the galactic frame) with the minimum and the maximum days set in the \verb|maddm_card.inc| file. The output file is written in \verb|output/VE_test.dat|. 
\item \verb|Form_test(Er, num_array):| Outputs the spin-dependent and spin-independent form factors for the most abundant elements. It uses the subroutine \verb|target_MA|, which is exactly the same as the routine \verb|target_material| above, but only provides the most abundant elements. When \verb|num_array| is \verb|0|, the output is the form factor at a specific input recoil energy. When \verb|num_array| is \verb|1|, the output is an array of the form factor squared in the file \verb|output/formfac_test.dat|. The first column is the recoil energy, the second is the spin-independent form factor squared while the third is the spin-dependent form factor squared.

\item \verb|Form_testmass(m_DM, num_array):| Outputs the spin-dependent and spin-independent form factors for the most abundant elements. It uses the subroutine \verb|target_MA|. When \verb|num_array| is \verb|0|, the output is the form factor at a specific input WIMP mass in GeV. When \verb|num_array| is \verb|1|, the output is an array of the form factor squared in the file \verb|output/formfacmass_test.dat|. The first column is the WIMP mass, the second is the spin-independent form factor squared while the third is the spin-dependent form factor squared.

\item \verb|LUX_Exclusion(xmim, xmax, ymin, ymax, step):| Scans the DM-nucleon vs Mass parameter space to provide the LUX exclusion limit. This is done by using the \verb|flag = 2| in the \verb|directional_detection| routine, which ensures that we use the efficiency function obtained from the LUX 2013 results \cite{Akerib:2013tjd}. \verb|xmin| represents the minimum mass, \verb|xmax|, the maximum mass, \verb|ymin| is the minimum cross-section and \verb|ymax| is the maximum cross-section in the scanning parameter range. \verb|nstep| is the number of scanning points for each direction. The scanning is done in log space, the linear scanning is commented out, for linear scanning, one has just to uncomment the \verb|Lux_exclusion| line in the test routine part of  \verb|maddm.f|
\end{itemize}



\bibliography{refs}
\end{document}